\documentclass[journal]{IEEEtran}
\usepackage{amsmath,amsfonts}
\usepackage{algorithm}
\usepackage{array}
\usepackage[caption=false,font=footnotesize,labelfont=rm,textfont=rm]{subfig}
\usepackage{textcomp}
\usepackage{stfloats}
\usepackage{url}
\usepackage{verbatim}
\usepackage{graphicx}
\usepackage{cite}
\usepackage{makecell}
\usepackage{algpseudocode}
\usepackage{multirow}
\usepackage{color, xcolor}
\usepackage{microtype}
\usepackage{amssymb}
\usepackage{balance}
\usepackage{amsthm}

\newcommand{\myvar}[1]{%
\lowercase{\def\tmp{#1}}%
\ifx\tmp x#1\else\MakeUppercase{#1}\fi}
\usepackage[colorlinks = true,
linkcolor = blue,
urlcolor  = blue,
citecolor = blue,
anchorcolor = blue]{hyperref}

\begin{document}

\title{Artificial-Noise-Aided Secure Near-Field MIMO With Fluid Antenna Systems}
\author{Peng~Zhang,~\IEEEmembership{Graduate~Student~Member,~IEEE,} Jian~Dang,~\IEEEmembership{Senior~Member,~IEEE,} Miaowen~Wen,~\IEEEmembership{Senior~Member,~IEEE,} Ziyang~Liu, Chen~Zhao,~\IEEEmembership{Senior~Member,~IEEE,} Huaifeng~Shi, Chengsheng~Pan, and Zaichen~Zhang,~\IEEEmembership{Senior~Member,~IEEE}


	\thanks{



		Peng Zhang, Jian Dang, and Zaichen Zhang are with the National Mobile Communications Research Laboratory and the Frontiers Science Center for Mobile Information Communication and Security, Southeast University, Nanjing 211189, China. Jian Dang is also with the Key Laboratory of Intelligent Support Technology for Complex Environments, Ministry of Education, Nanjing University of Information Science and Technology, Nanjing 210044, China. All three authors are also with Purple Mountain Laboratories, Nanjing 211111, China (e-mail: \mbox{peng\_zhang@seu.edu.cn}; \mbox{dangjian@seu.edu.cn}; \mbox{zczhang@seu.edu.cn}).

Miaowen Wen is with the School of Electronic and Information Engineering, South China University of Technology, Guangzhou 510640, China (e-mail: \mbox{eemwwen@scut.edu.cn}).

Ziyang Liu is with the School of Communication Engineering, Hangzhou Dianzi University, Hangzhou 310018, China (e-mail: \mbox{251080010@hdu.edu.cn}).

Chen Zhao and Huaifeng Shi are with the School of Electronics and Information Engineering, Nanjing University of Information Science and Technology, Nanjing 210044, China (e-mail: \mbox{002912@nuist.edu.cn}; \mbox{shihuaifeng@nuist.edu.cn}).

Chengsheng Pan is with the School of Communication and Information Engineering, Nanjing University of Posts and Telecommunications, Nanjing 210003, China (e-mail: \mbox{pancs@nuist.edu.cn}).

Corresponding authors: Jian Dang and Zaichen Zhang (e-mail: \mbox{dangjian@seu.edu.cn}; \mbox{zczhang@seu.edu.cn}).

	}}

\markboth{IEEE Transactions on Mobile Computing,~Vol.~0, No.~0, August~2025}%
{Shell \MakeLowercase{\textit{et al.}}: A Sample Article Using IEEEtran.cls for IEEE Journals}


\maketitle

\begin{abstract}

	With the evolution of mobile communication systems toward large-scale arrays, high-frequency operation, and reconfigurable antenna architectures, fluid antenna systems (FAS) operating in the near-field (NF) regime provide new degrees of freedom (DoF) for secure and privacy-sensitive mobile access. This paper proposes an artificial-noise (AN)-aided physical layer security (PLS) scheme for NF fluid-antenna multiple-input multiple-output (FA-MIMO) systems, aiming to protect high-rate mobile service links supported by compact or large arrays. An alternating-optimization (AO) framework addresses the sparsity-constrained non-convex design by splitting it into a continuous BF/AN joint-design subproblem and a discrete FAS port-selection subproblem. Closed-form fully digital beamforming (BF)/AN solutions are obtained via a generalized spectral water-filling procedure within a block coordinate descent (BCD) surrogate and realized by a hardware-consistent hybrid beamforming (HBF) architecture with a shared RF network and independent digital BF/AN branches, while preserving the target BF/AN power split under constant-modulus RF constraints. For FAS port selection, a row-energy based prune--refit rule, aligned with Karush--Kuhn--Tucker (KKT) conditions of a group-sparsity surrogate, enables efficient active-port determination under a finite RF-chain budget. Simulation results confirm that the proposed design exploits the geometry and position-domain DoF of FAS and significantly improves secrecy performance, particularly for non-extremely-large arrays where NF beam focusing alone is inadequate. These results demonstrate the potential of AN-aided NF FA-MIMO as a practical secure-transmission architecture for future location-aware and hardware-constrained mobile computing systems.

\end{abstract}


\section{Introduction}
\label{Introduction}

\IEEEPARstart{A}{s} mobile computing evolves toward the sixth-generation (6G) era, intelligent and delay-sensitive applications such as immersive extended reality (XR), autonomous driving, remote healthcare, and large-scale Internet of Things (IoT) are becoming pervasive and mission-critical \cite{Wang2023On}. These services require mobile devices to continuously exchange high-rate and privacy-sensitive service data with access points, edge servers, and surrounding intelligent terminals, including user identity, behavioral patterns, physiological parameters, location data, and environmental sensing signals. However, the openness and time-varying nature of wireless access links expose such mobile data to eavesdropping and information leakage risks \cite{Abuhamad2021Sensor}.

As an important component of mobile information security, physical-layer security (PLS) protects privacy-sensitive service data over wireless access links by exploiting the quality disparity between legitimate and eavesdropping channels, which is commonly measured by secrecy rate (SR) and secrecy outage probability \cite{Leung-Yan-Cheong1978The,Nguyen2025Reliable,Soderi2024MultiRIS,Li2026MA_Mobility}. In mobile secure access, this disparity can be enhanced through cooperative relaying, friendly jamming, user scheduling, and radio-resource allocation \cite{Yue2025Power,He2019Joint}. At the waveform level, multiantenna beamforming (BF) strengthens the legitimate link and suppresses leakage by antenna-weight design, while statistical or location-aware BF preserves spatial selectivity under partial channel state information (CSI) \cite{Zhao2025Joint}. Artificial noise (AN) further degrades unauthorized decoding by injecting interference that is weakly visible to the legitimate receiver but harmful to the eavesdropper \cite{Niu2025Survey}. Joint information--AN design with power allocation balances main-link enhancement and wiretap-link degradation \cite{shi2015Secure,Fang2022Intelligent}, and can be extended to multiuser and multicarrier mobile networks through user/subcarrier coordination \cite{Yang2020Artificial,Yi2025Secrecy}. Robust designs under imperfect CSI and unobservable passive eavesdroppers are also needed to sustain secrecy in mobile environments \cite{Luo2024Joint}. For the considered location-aware secure transmission design, the locations of both the legitimate user and the potential eavesdropper are assumed to be obtainable through positioning, integrated sensing and communication, or location-aware channel acquisition techniques exploiting geographical and angular information~\cite{wang2024SensingCovert,Xia2026ISACBeamforming,Zhang2026JADCE}.

With the emergence of massive multiple-input multiple-output (mMIMO), millimeter-wave (mmWave), and terahertz (THz) technologies, the Rayleigh distance of antenna arrays has been significantly extended due to enlarged apertures and reduced wavelengths \cite{lu2023Near}. As a result, practical mobile access links may operate in the near-field (NF) region, where electromagnetic (EM) propagation is characterized by spherical wavefronts and spatially nonuniform amplitude and phase responses depending on both angle and distance \cite{Wu2024Enabling}. These angle--distance-dependent channel characteristics provide additional spatial selectivity for PLS, which is particularly useful for protecting location-sensitive and privacy-critical mobile services. Existing NF-PLS studies have exploited this property from different perspectives. Directional modulation with fully analog precoding and embedded AN was adopted in \cite{Chen2024NFDM} to secure transmission in both angle and distance domains. A max--min secrecy beam-focusing problem under a hybrid beamforming (HBF) architecture was formulated in \cite{Nasir2024MaxMin} to improve the worst-case secrecy performance. The work in \cite{Zhang2025NFPLS} analyzed near-field secrecy performance and developed a low-complexity beamforming design, showing that structured AN can approach near-optimal secrecy performance. For THz systems, \cite{Tang2025ANBF} steered the information and AN beams toward the legitimate user and the eavesdropper, respectively, and derived a closed-form power-allocation solution. In addition, wavefront hopping with engineered Bessel and Airy profiles was investigated in \cite{Petrov2025WFH} to reshape the NF field distribution and reduce interception probability. Nevertheless, the above designs are mainly restricted to multiple-input single-output (MISO) links. In NF-MIMO systems, the effective apertures at both the transmitter and receiver enlarge the Rayleigh distance and introduce stronger angle--distance coupling across multiple spatial streams, which makes secure mobile access more dependent on joint spatial, power, and hardware-aware signal design. A recent study \cite{zhang2024physical} extended NF-PLS to a MIMO setting and maximized secrecy through a two-stage HBF design under spherical-wave CSI, where secrecy mainly relies on beam focusing. However, focusing-dominant designs are most effective for very large arrays, whereas compact or non-extremely-large mmWave/THz MIMO arrays may still operate in the NF region but provide limited focusing gain. This motivates a joint BF--AN design for NF-MIMO and the exploration of additional antenna-domain DoF, so as to support secure and hardware-constrained mobile communication systems.

Fluid antenna (FA) technology \cite{Zhang2025UNISAC,Zhang2025FARDIM,Zhang2025SF_FAMA} enables dynamic repositioning of the radiating element or active port within a compact region, thereby creating position-domain diversity on top of conventional spatial and pattern diversity and offering a flexible means to combat fading, interference, and eavesdropping with compact hardware. Recent studies have started to exploit this capability for secure and covert communications. In \cite{FAS_Secure_Covert}, FA-enabled transceivers are used to enhance secrecy and covertness by opportunistically selecting favorable antenna positions. Continuous-trajectory FA index modulation \cite{Zhang2024RDRSM} is introduced in \cite{Liu2025IM_CT_FAS} to embed covert information into the antenna trajectory, while \cite{FAS_Secrecy_Analysis} provides secrecy performance analysis of FA-assisted wiretap channels under spatial correlation. The integration of reconfigurable intelligent surfaces (RIS) with FA is examined in \cite{RIS_FAS_Secrecy}, showing additional gains from joint reflection and position diversity, and \cite{CUMA_Compact_Ultra_Massive} demonstrates that compact ultra-massive FA arrays can support reliable and secure multiuser transmission via port selection and interference shaping. Furthermore, trajectory- and geometry-aware FA designs are investigated in \cite{Trajectory_FA_Secure,Chen2025FA_3D_Covert_Jamming}, where mechanical or three-dimensional FA movement is optimized and combined with friendly jamming to reinforce secure or covert links. Recent studies have also investigated geometry-aware fluid antenna array design under finite-aperture constraints, establishing analytical performance limits and practical placement algorithms \cite{Zhang2026FiniteFA}. These works collectively confirm that FA can effectively enhance PLS by enriching the available DoF through spatial and positional reconfiguration.

However, most existing FA-based secure schemes are formulated under far-field (FF) or simplified propagation models and predominantly employ BF, port selection, or trajectory control in single-antenna or MISO-type settings. They do not explicitly address high-frequency MIMO systems with non-extremely-large arrays, where the Rayleigh distance is still comparable to practical link distances and near-field characteristics such as angle--distance coupling and distance-selective focusing remain pronounced. In this regime, secure transmission becomes jointly shaped by propagation geometry, RF-chain constraints, and port-position reconfiguration, so BF-only FA designs provide limited capability to exploit the full NF structure for secrecy enhancement. Motivated by these observations, this work studies AN-aided secure NF-MIMO mobile-access transmission with FAS under a finite RF-chain budget. The proposed design emphasizes the joint effect of NF spherical-wave propagation, FA port-position reconfiguration, AN injection, and hybrid realization, so that the physical port locations and the continuous BF/AN variables are optimized in a unified hardware-constrained secure mobile-access procedure.

To summarize, the
work and contributions of this paper are as follows:
\begin{itemize}
	\item We develop an AN-aided NF fluid-antenna MIMO (FA-MIMO) secure mobile access system with a discretized fluid antenna system (FAS) transmitter. Based on this system, we formulate a joint BF/AN design and active-port selection problem for maximizing the SR under a finite radio-frequency (RF)-chain budget.

	\item For a fixed active-port set, we derive a structured fully digital BF/AN update by decomposing the SR into tractable log-det terms and applying block coordinate descent (BCD). The resulting subproblem admits closed-form updates through generalized spectral-domain water filling and a scalar water-level search.

	\item For active-port selection and implementation, we develop a row-energy based progressive prune--refit rule linked to the Karush--Kuhn--Tucker (KKT) structure of a group-sparsity surrogate. This rule moves RF-connected ports toward high-utility regions of the FA rail, and the resulting fully digital target is then mapped to a shared-RF HBF architecture after port selection.
\end{itemize}

The rest of this paper is organized as follows.
Section~\ref{section_system_model} introduces the NF FA-MIMO secure mobile-access system, including the near-field channel model, the signal model, and the corresponding secrecy-rate maximization problem.
Section~\ref{section_BF_AN_optimization} presents the proposed secure mobile-access design, including the fully digital BF/AN optimization for a fixed active-port set, the spectral-domain power-allocation and stream-balancing strategy, the progressive active-port selection procedure, the hybrid implementation over the selected ports, and the overall algorithm with complexity analysis.
Section~\ref{sec:numerical_results} provides numerical results and discussions.
Finally, Section~\ref{conclusion} concludes the paper.

\emph{Notations:} Bold lowercase and uppercase letters denote vectors and matrices, respectively. The operators \((\cdot)^{\mathrm T}\) and \((\cdot)^{\mathrm H}\) denote transpose and conjugate transpose, respectively. The imaginary unit is denoted by \(\iota=\sqrt{-1}\), and \(\mathbf I_n\) denotes the \(n\times n\) identity matrix. The operators \(\operatorname{tr}(\cdot)\), \(\det(\cdot)\), \(\operatorname{diag}(\cdot)\), \(|\cdot|\), and \(\angle(\cdot)\) denote the trace, determinant, diagonal formation/extraction, scalar magnitude, and elementwise phase, respectively; \(\Re\{\cdot\}\) and \(\Im\{\cdot\}\) denote the real and imaginary parts. The sparsity measure, Euclidean norm, and Frobenius norm are denoted by \(\|\cdot\|_0\), \(\|\cdot\|_2\), and \(\|\cdot\|_{\mathrm F}\), respectively. The expectation operator is denoted by \(\mathbb E\{\cdot\}\). A circularly symmetric complex Gaussian random variable with zero mean and variance \(\sigma^2\) is represented by \(\mathcal{CN}(0,\sigma^2)\). The sets \(\mathbb C^{m\times n}\) and \(\mathbb B^{m\times n}\) denote complex-valued and binary-valued \(m\times n\) matrices, respectively, and \(\mathbb B^n\) denotes binary vectors of length \(n\). Calligraphic symbols such as \(\mathcal S\) denote index sets, with \(\mathbf P_{\mathcal S}\) denoting the associated column-selection matrix. For a matrix \(\mathbf X\), \(\mathbf X_{i,:}\), \(\mathbf X_{:,j}\), and \([\mathbf X]_{i,j}\) denote its \(i\)-th row, \(j\)-th column, and \((i,j)\)-th entry, respectively. For notational convenience, \(\mathbf x_n \triangleq (\mathbf X_{n,:})^{\mathrm H}\) denotes the column vector obtained from the \(n\)-th row of \(\mathbf X\) by conjugate transposition. For a real scalar \(x\), \([x]^+\triangleq \max\{x,0\}\) denotes its nonnegative part. The relations \(\mathbf A\succeq \mathbf 0\) and \(\mathbf A\succ \mathbf 0\) mean that \(\mathbf A\) is Hermitian positive semidefinite and positive definite, respectively.

\section{System Model}
\label{section_system_model}
\begin{figure}[!t]
	\centering
	\includegraphics[width=3.4in]{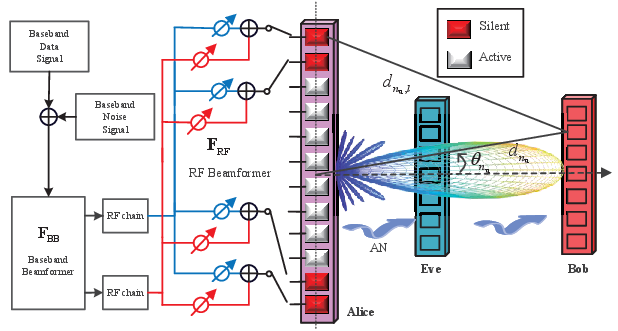}
	\caption{Illustration of FA port array at the transmitter and FPA array at the receivers.\label{antenna_system}}
\end{figure}

As shown in Fig.~\ref{antenna_system}, a FAS-enabled NF-MIMO secure mobile-access system consists of a transmitter (Alice), a legitimate mobile service endpoint (Bob), and a potential eavesdropper (Eve). At the transmitter, the FAS comprises $L$ densely deployed candidate FA ports, among which $N_{\text{t}}$ ports are activated and connected to the RF chains, with $L \ge N_{\text{t}} \gg 1$. We adopt a discretized circuit-driven FAS architecture that is suitable for practical hardware implementation, where the active ports are selected electronically through a reconfigurable switching network rather than by liquid or mechanical motion. This abstraction is consistent with electronically reconfigurable FAS realizations, including switch-controlled implementations such as pixel-based FAS~\cite{zhang2024PRA}. For comparison, Bob and Eve employ conventional FPA arrays with $N_{\text{u}}$ and $N_{\text{e}}$ uniformly spaced elements, respectively, where $N_{\text{t}} \gg N_{\text{u}} \ge 1$ and $N_{\text{t}} \gg N_{\text{e}} \ge 1$.

In MIMO systems operating at high-frequency bands, the near-field characteristics of electromagnetic propagation become significant.
The corresponding Rayleigh distance is expressed as $r_{\text{MIMO-RD}} = \tfrac{2\left(D_{\text{A}} + D_{\text{B/E}}\right)^{2}}{\lambda}$~\cite{lu2023Near},
where $\lambda$ is the carrier wavelength, $D_{\text{A}}$, $D_{\text{B}}$, and $D_{\text{E}}$ denote the physical apertures of the antenna arrays at Alice, Bob, and Eve, respectively.
CSI acquisition is generally difficult in mobile access systems and becomes even more challenging when the channel of a potential eavesdropper is involved. In this work, we adopt a location-aware secure mobile-access setting motivated by privacy-sensitive mobile services. In particular, a potential eavesdropper may appear as a line-of-sight (LoS) terminal in the service area, such as a household robot or other connected device whose location can be inferred or tracked. In such cases, the transmitter can perform directional shielding and jamming toward the corresponding spatial direction and range. Since near-field propagation is highly sensitive to the spatial positions of the transceivers, the locations of both the legitimate user and the potential eavesdropper are assumed to be available through positioning or integrated sensing and communication techniques~\cite{Su2024Sensing,Hou2024Optimal}. Accordingly, the CSIs of the Alice--Bob and Alice--Eve LoS links are regarded as known for secure transmission design.

\subsection{Near-Field Fluid Antenna Channel Model}
\begin{figure}[!t]
	\centering
	\includegraphics[width=3.4in]{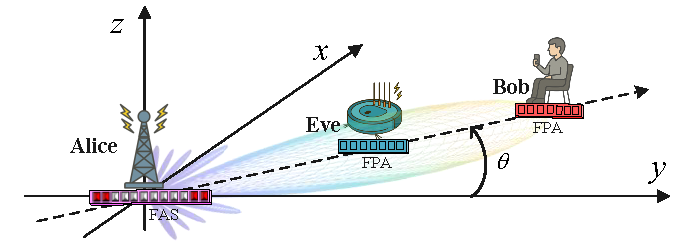}
	\caption{Geometry of the near-field FA-MIMO wiretap system with the Cartesian coordinate setup.\label{coordination_figure}}
\end{figure}

As illustrated in Fig.~\ref{coordination_figure}, a Cartesian coordinate system is established with the array center of the transmitter located at the origin $(0,0,0)$.
This work considers linearly arranged FAs.
The arrays of Alice, Bob, and Eve are all parallel to the \(y\)-axis, and both receivers are positioned at the same height as the transmitter.
The position of Alice's \(l\)-th port is \((0,\Delta y_l,0)\), where
\(\Delta y_l=\big(l-\frac{L+1}{2}\big)d_{\text{port}}\).
The centers of the FPA arrays at Bob and Eve are located at
\((x_{\text{B}},y_{\text{B}},z_{\text{B}})=(d_{\text{B}}\cos\theta_{\text{B}},d_{\text{B}}\sin\theta_{\text{B}},0)\)
and
\((x_{\text{E}},y_{\text{E}},z_{\text{E}})=(d_{\text{E}}\cos\theta_{\text{E}},d_{\text{E}}\sin\theta_{\text{E}},0)\), respectively.
Here, \(d_{\text{B}}\) and \(d_{\text{E}}\) denote the radial distances from Alice to Bob and Eve, and \(\theta_{\text{B}}\in[0,\pi]\) and \(\theta_{\text{E}}\in[0,\pi]\) are their azimuth angles.
The position of Bob's \(n_{\text{u}}\)-th antenna element is
\((x_{\text{B}},\,y_{\text{B}}+\Delta y_{n_{\text{u}}},0)\), and the position of Eve's \(n_{\text{e}}\)-th antenna element is
\((x_{\text{E}},\,y_{\text{E}}+\Delta y_{n_{\text{e}}},0)\).
The offsets are defined as
\(\Delta y_{n_{\text{u}}}=\big(n_{\text{u}}-\frac{N_{\text{u}}+1}{2}\big)d_{\text{FPA}}\)
and
\(\Delta y_{n_{\text{e}}}=\big(n_{\text{e}}-\frac{N_{\text{e}}+1}{2}\big)d_{\text{FPA}}\),
where \(d_{\text{FPA}}\) denotes the element spacing of the FPA arrays for both Bob and Eve.
To investigate the impact of massive MIMO on physical layer security in the distance domain under near-field conditions, we focus on the case $\theta_{\text{B}}=\theta_{\text{E}}=\theta$.
This configuration represents a technically most challenging scenario.

In this system, the exact distances can be written in closed form.
The distance from the $l$-th FA port at Alice to the $n_{\text{u}}$-th receive antenna element at Bob is
\begin{equation}
	\label{distance_AB_exact}
	d_{n_{\text{u}},l} = \sqrt{\, x_{\text{B}}^2 + \big(y_{\text{B}} + \Delta y_{n_{\text{u}}} - \Delta y_{l}\big)^2 \,},
\end{equation}
and the distance from the $n_{\text{u}}$-th receive antenna to the array reference point at Alice is
\begin{equation}
	\label{distance_AB_ref}
	d_{n_{\text{u}}} = \sqrt{\, x_{\text{B}}^2 + \big(y_{\text{B}} + \Delta y_{n_{\text{u}}}\big)^2 \,},
\end{equation}
where $x_{\text{B}} = d_{\text{B}}\cos\theta_{\text{B}}$ and $y_{\text{B}} = d_{\text{B}}\sin\theta_{\text{B}}$.

For analytical tractability under near-field conditions, the Fresnel approximation is applied when $d_{\text{B}} \gg d_{\text{FPA}}, d_{\text{port}}$.
The first-order Taylor expansion of the distance function can be expressed as
\begin{equation}
	\label{taylor_expansion}
	\sqrt{d_{\text{B}}^2 + \varepsilon} \;\approx\; d_{\text{B}} + \frac{\varepsilon}{2d_{\text{B}}}, \quad |\varepsilon|\ll d_{\text{B}}^2,
\end{equation}
which leads to the approximated distances
\begin{equation}
	\label{distance_AB_fresnel}
	d_{n_{\text{u}},l} \approx d_{\text{B}} + \sin\theta_{\text{B}}\big(\Delta y_{n_{\text{u}}} - \Delta y_{l}\big) + \frac{(\Delta y_{n_{\text{u}}} - \Delta y_{l})^2}{2d_{\text{B}}},
\end{equation}
\begin{equation}
	\label{distance_AB_ref_fresnel}
	d_{n_{\text{u}}} \approx d_{\text{B}} + \sin\theta_{\text{B}}\Delta y_{n_{\text{u}}} + \frac{(\Delta y_{n_{\text{u}}})^2}{2d_{\text{B}}}.
\end{equation}

The derived results provide both the exact and approximate distance formulations based on this Taylor approximation.
These formulations characterize the geometric propagation in the near-field region and serve as the basis for computing the corresponding phase variations in the channel model.
Following the same derivation procedure, the channel between the transmitter and the potential eavesdropper can be obtained by replacing Bob’s antenna coordinates and array size $N_{\text{u}}$ with those of Eve, that is, $(x_{\text{E}}, y_{\text{E}})$ and $N_{\text{e}}$.

Based on the above geometry, the LoS channel from Alice to Bob is defined as
\begin{equation}
\begin{aligned}
	\relax[\mathbf{H}_{\text{B},\mathrm{LoS}}]_{n_{\text{u}},l}
	=
	\frac{c}{4\pi f d_{n_{\text{u}},l}}
	\exp\!\left(
	-\iota\frac{2\pi f}{c}(d_{n_{\text{u}},l}-d_{n_{\text{u}}})
	\right),
\end{aligned}
\label{channel_model_AB_LoS}
\end{equation}
for $n_{\text{u}}=1,\ldots,N_{\text{u}}$ and $l=1,\ldots,L$. Similarly, the LoS channel from Alice to Eve is
\begin{equation}
\begin{aligned}
	\relax[\mathbf{H}_{\text{E},\mathrm{LoS}}]_{n_{\text{e}},l}
	=
	\frac{c}{4\pi f d_{n_{\text{e}},l}}
	\exp\!\left(
	-\iota\frac{2\pi f}{c}(d_{n_{\text{e}},l}-d_{n_{\text{e}}})
	\right),
\end{aligned}
\label{channel_model_AE_LoS}
\end{equation}
for $n_{\text{e}}=1,\ldots,N_{\text{e}}$ and $l=1,\ldots,L$.

To account for the diffuse scattering in practical propagation environments, the non-LoS (NLoS) channels are modeled as
\begin{equation}
\begin{aligned}
	\relax[\mathbf{H}_{\text{B},\mathrm{NLoS}}]_{n_{\text{u}},l}
	=
	\frac{c}{4\pi f d_{n_{\text{u}},l}}\,w_{\text{B},n_{\text{u}},l},
\end{aligned}
\label{channel_model_AB_NLoS}
\end{equation}
\begin{equation}
\begin{aligned}
	\relax[\mathbf{H}_{\text{E},\mathrm{NLoS}}]_{n_{\text{e}},l}
	=
	\frac{c}{4\pi f d_{n_{\text{e}},l}}\,w_{\text{E},n_{\text{e}},l},
\end{aligned}
\label{channel_model_AE_NLoS}
\end{equation}
where $w_{\text{B},n_{\text{u}},l}$ and $w_{\text{E},n_{\text{e}},l}$ are i.i.d. complex Gaussian random variables following $\mathcal{CN}(0,1)$.

Accordingly, we adopt a correlated Rician near-field channel model. The Alice--Bob and Alice--Eve channels are respectively given by
\begin{equation}
\begin{aligned}
	\mathbf{H}_{\text{B}}
	=
	\frac{1}{\sqrt{N_{\text{t}}}}
	\Big(
	\sqrt{\frac{K_{\text{B}}}{K_{\text{B}}+1}}\mathbf{H}_{\text{B},\mathrm{LoS}}
	+
	\sqrt{\frac{1}{K_{\text{B}}+1}}\mathbf{H}_{\text{B},\mathrm{NLoS}}
	\Big),
\end{aligned}
\label{channel_model_AB}
\end{equation}
\begin{equation}
\begin{aligned}
	\mathbf{H}_{\text{E}}
	=
	\frac{1}{\sqrt{N_{\text{t}}}}
	\Big(
	\sqrt{\frac{K_{\text{E}}}{K_{\text{E}}+1}}\mathbf{H}_{\text{E},\mathrm{LoS}}
	+
	\sqrt{\frac{1}{K_{\text{E}}+1}}\mathbf{H}_{\text{E},\mathrm{NLoS}}
	\Big),
\end{aligned}
\label{channel_model_AE}
\end{equation}
with $K_{\text{B}}$ and $K_{\text{E}}$ denoting the Rician factors of the Alice--Bob and Alice--Eve links, respectively.

In addition, to characterize the correlation among densely deployed candidate FA ports caused by dense spacing and mutual coupling, we introduce an effective transmit-side spatial correlation matrix $\mathbf{J}\in\mathbb{C}^{L\times L}$, whose $(l,m)$-th entry is modeled as
\begin{equation}
	[\mathbf{J}]_{l,m}
	=
	J_0\!\left(
	\frac{2\pi|\Delta y_l-\Delta y_m|}{\lambda}
	\right),
\label{corr_matrix}
\end{equation}
where $J_0(\cdot)$ is the zeroth-order Bessel function of the first kind. Since $\mathbf J$ is a Hermitian positive semidefinite correlation matrix, its eigendecomposition can be written as
\begin{equation}
	\mathbf J=\mathbf U_{\mathbf J}\mathbf\Lambda_{\mathbf J}\mathbf U_{\mathbf J}^{\mathrm H},
\end{equation}
where $\mathbf U_{\mathbf J}$ contains the eigenvectors of $\mathbf J$, and $\mathbf\Lambda_{\mathbf J}=\operatorname{diag}(\lambda_1,\lambda_2,\ldots,\lambda_L)$ collects the corresponding nonnegative eigenvalues. Accordingly, the Hermitian square root of $\mathbf J$ is given by
\begin{equation}
	\mathbf J^{1/2}
	=
	\mathbf U_{\mathbf J}\mathbf\Lambda_{\mathbf J}^{1/2}\mathbf U_{\mathbf J}^{\mathrm H},
\end{equation}
with $\mathbf\Lambda_{\mathbf J}^{1/2}=\operatorname{diag}(\sqrt{\lambda_1},\sqrt{\lambda_2},\ldots,\sqrt{\lambda_L})$, which satisfies $\mathbf J^{1/2}(\mathbf J^{1/2})^{\mathrm H}=\mathbf J$.

It is worth noting that when $K_{\text{B}},K_{\text{E}}\rightarrow\infty$, the above model reduces to the original deterministic near-field LoS channel model as a special case.

\subsection{Secure Transmission Signal Model}
In the considered secure transmission system, Alice sends confidential information to Bob in the presence of a potential eavesdropper, Eve.
We consider the challenging wiretap scenario where Eve is closer to Alice than Bob, i.e., $d_{\mathrm E}<d_{\mathrm B}$.
To improve the secrecy performance, Alice adopts an HBF architecture equipped with an FA array, where only $N_{\mathrm t}$ out of the $L$ densely deployed candidate FA ports are activated and connected to the RF chains.

Let $\mathcal S=\{s_1,\ldots,s_{|\mathcal S|}\}\subseteq\{1,\ldots,L\}$ denote the active-port support at a given selection stage.
The associated selection matrix is defined as
$\mathbf P_{\mathcal S}\triangleq[\mathbf e_{s_1},\ldots,\mathbf e_{s_{|\mathcal S|}}]\in\mathbb B^{L\times|\mathcal S|}$,
where $\mathbf e_l$ denotes the $l$-th column of the $L\times L$ identity matrix.
Accordingly, $\mathbf P_{\mathcal S}$ maps reduced-dimensional port-domain variables defined over the active support into the original $L$-port candidate space, while $\bar{\mathbf H}\mathbf P_{\mathcal S}$ extracts the channel columns indexed by $\mathcal S$ from a full channel matrix $\bar{\mathbf H}$.
It follows that $\mathbf P_{\mathcal S}^{\mathrm T}\mathbf P_{\mathcal S}=\mathbf I_{|\mathcal S|}$, and each row of $\mathbf P_{\mathcal S}$ contains at most one nonzero entry.
After the pruning--refitting procedure converges, the terminal support set $\mathcal S^\star$ satisfies $|\mathcal S^\star|=N_{\mathrm t}$, and the corresponding selection matrix becomes $\mathbf P_{\mathcal S^\star}\in\mathbb B^{L\times N_{\mathrm t}}$.

Let the dense HBF targets be $\tilde{\mathbf{W}}\in\mathbb{C}^{N_{\text t}\times K}$ for the $K$-stream service-data vector $\mathbf{x}\in\mathbb{C}^{K\times 1}$ and $\tilde{\mathbf v}\in\mathbb{C}^{N_{\text t}\times 1}$ for AN.
Row selection on the activated ports yields the sparse FA–MIMO BF matrix and AN vector as
\begin{equation}
	\label{eq:fa_bf_selection_an}
	\mathbf{W}=\mathbf{P}_{\mathcal S^\star}\tilde{\mathbf{W}},
\end{equation}
\begin{equation}
	\label{eq:fa_selection_an}
	\mathbf{v}=\mathbf{P}_{\mathcal S^\star}\tilde{\mathbf v}.
\end{equation}
Here $\tilde{\mathbf v}\in\mathbb C^{N_{\text t}\times 1}$ denotes the dense AN vector before row selection.

Let $z\sim\mathcal{CN}(0,1)$ be a scalar AN symbol independent of $\mathbf{x}$. The transmit vector is
\begin{equation}
	\label{transmit_signal}
	\mathbf{s}
	=\mathbf{P}_{\mathcal S^\star}\big(\tilde{\mathbf{W}}\mathbf x+\tilde{\mathbf v}\,z\big)
	=\mathbf W\mathbf x+\mathbf v\,z
	\ \in\ \mathbb{C}^{L\times 1}.
\end{equation}

The total transmit power satisfies
\begin{equation}
	\label{power_constraint}
	\operatorname{Tr}(\mathbf{W}\mathbf{W}^{\mathrm H})
	+\operatorname{Tr}(\mathbf{v}\mathbf{v}^{\mathrm H})
	\;\le\; P_{\mathrm t}.
\end{equation}

The received signals at Bob and Eve are
\begin{equation}
	\label{received_signal_B}
	\mathbf{y}_{\text{\text{B}}}=\mathbf{H}_{\text{\text{B}}}\mathbf{J}^{1/2}\mathbf{s}+\mathbf{n}_{\text{\text{B}}}
	=\mathbf{H}_{\text{\text{B}}}\mathbf{J}^{1/2}\mathbf{W}\mathbf{x}+\mathbf{H}_{\text{\text{B}}}\mathbf{J}^{1/2}\mathbf{v}\,z+\mathbf{n}_{\text{\text{B}}},
\end{equation}
\begin{equation}
	\label{received_signal_E}
	\mathbf{y}_{\text{E}}=\mathbf{H}_{\text{E}}\mathbf{J}^{1/2}\mathbf{s}+\mathbf{n}_{\text{E}}
	=\mathbf{H}_{\text{E}}\mathbf{J}^{1/2}\mathbf{W}\mathbf{x}+\mathbf{H}_{\text{E}}\mathbf{J}^{1/2}\mathbf{v}\,z+\mathbf{n}_{\text{E}},
\end{equation}
where $\mathbf{n}_{\text{\text{B}}}\sim\mathcal{CN}(0,\sigma^2\mathbf I_{N_{\text u}})$ and
$\mathbf{n}_{\text{E}}\sim\mathcal{CN}(0,\sigma^2\mathbf I_{N_{\text e}})$ are additive white Gaussian noise (AWGN).
To simplify the subsequent BF/AN optimization, we absorb the transmit correlation directly into the noise-whitened equivalent channels and define
\begin{equation}
	\bar{\mathbf H}_{\text U}\triangleq \mathbf{H}_{\text{\text{B}}}\mathbf J^{1/2}/\sqrt{\sigma^2},\qquad
	\bar{\mathbf H}_{\text E}\triangleq \mathbf{H}_{\text{E}}\mathbf J^{1/2}/\sqrt{\sigma^2}.
\end{equation}
Defining the signal and AN covariances
\begin{equation}
	\mathbf S\triangleq \mathbf W\mathbf W^{\text H},\qquad
	\mathbf A_{\rm AN}\triangleq \mathbf v\mathbf v^{\text H},
\end{equation}
the achievable data rates are
\begin{equation}
	\label{data_rate_B}
	R_{\text{\text{B}}}
	=\log\det\!\Big(
	\mathbf I_{N_{\text u}}
	+\bar{\mathbf H}_{\text U}\,\mathbf S\,\bar{\mathbf H}_{\text U}^{\text H}\,
	(\mathbf I_{N_{\text u}}+\bar{\mathbf H}_{\text U}\,\mathbf A_{\rm AN}\,\bar{\mathbf H}_{\text U}^{\text H})^{-1}
	\Big),
\end{equation}
\begin{equation}
	\label{data_rate_E}
	R_{\text{E}}
	=\log\det\!\Big(
	\mathbf I_{N_{\text e}}
	+\bar{\mathbf H}_{\text E}\,\mathbf S\,\bar{\mathbf H}_{\text E}^{\text H}\,
	(\mathbf I_{N_{\text e}}+\bar{\mathbf H}_{\text E}\,\mathbf A_{\rm AN}\,\bar{\mathbf H}_{\text E}^{\text H})^{-1}
	\Big),
\end{equation}
and the SR is
\begin{equation}
	\label{secrecy_rate}
	R_{\text{s}}=\big[R_{\text{\text{B}}}-R_{\text{E}}\big]^+.
\end{equation}

In the imperfect-CSI case, Bob's CSI is assumed to be perfectly known, whereas the CSI of Eve used for transmission design is imperfect. Specifically, the estimated Eve channel is modeled as
\begin{equation}
\label{eq:imperfect_eve_csi}
\widehat{\mathbf H}_{\mathrm E}
=
\mathbf H_{\mathrm E}
+
\Delta_{\mathrm E},
\end{equation}
where \(\widehat{\mathbf H}_{\mathrm E}\) is the estimated Eve channel available at the transmitter, \(\mathbf H_{\mathrm E}\) is the true Eve channel, and \(\Delta_{\mathrm E}\) denotes the channel estimation error. The channel uncertainty level is characterized by the normalized mean-square error (NMSE), defined as \(\mathrm{NMSE}=\mathbb{E}\!\left[\|\Delta_{\mathrm E}\|_{\mathrm F}^{2}\right]/\mathbb{E}\!\left[\|\mathbf H_{\mathrm E}\|_{\mathrm F}^{2}\right]\). Accordingly, port selection, BF/AN design, and HBF realization are carried out based on \(\widehat{\mathbf H}_{\mathrm E}\), while the final secrecy rate is evaluated on the true channel \(\mathbf H_{\mathrm E}\).

\subsection{Secrecy Rate Maximization Problem}
The preceding signal model naturally gives rise to a joint SR maximization problem involving confidential BF, AN injection, and FA-port selection.
Instead of directly enumerating all possible active-port supports, we formulate the port-selection task through a joint row-sparsity constraint imposed on the effective port-domain BF matrix and AN vector.
Specifically, let $\mathbf W\in\mathbb C^{L\times K}$ and $\mathbf v\in\mathbb C^{L\times 1}$ denote the effective BF matrix and AN vector over the $L$ candidate FA ports, respectively.
Based on the secrecy rate definition in \eqref{secrecy_rate}, the joint design problem is formulated as

\begin{subequations}\label{eq:sr_sparse_problem_brief}
	\begin{align}
		\max_{\mathbf W,\,\mathbf v}\quad
		&
		R_{\mathrm s}\big(\mathbf W,\mathbf v\big),
		\label{eq:sr_sparse_problem_brief:a}
		\\
		\mathrm{s.t.}\quad
		&
		\operatorname{Tr}\big(
		\mathbf W^{\mathrm H}\mathbf W
		\big)
		+
		\big\|
		\mathbf v
		\big\|_2^2
		\le P_{\mathrm t},
		\label{eq:sr_sparse_problem_brief:b}
		\\
		&
		\left\|
		\left[
		\left\|
		\left[
		\mathbf W_{i,:},\,
		[\mathbf v]_{i}
		\right]
		\right\|_2
		\right]_{i=1}^{L}
		\right\|_0
		\le N_{\mathrm t},
		\label{eq:sr_sparse_problem_brief:c}
		\\
		&
		\mathbf W\in\mathbb C^{L\times K},
		\quad
		\mathbf v\in\mathbb C^{L\times 1}.
		\label{eq:sr_sparse_problem_brief:d}
	\end{align}
\end{subequations}

Here, Constraint~\eqref{eq:sr_sparse_problem_brief:b} limits the total transmit power allocated to the confidential data streams and AN. 
Constraint~\eqref{eq:sr_sparse_problem_brief:c} imposes a joint row-sparsity budget on $\mathbf W$ and $\mathbf v$, such that the number of FA ports carrying either data or AN components does not exceed $N_{\mathrm t}$.
The active-port support induced by the nonzero rows is given by

\begin{equation}
	\mathcal S^\star
	=
	\left\{
	i\in\{1,\ldots,L\}
	\;\middle|\;
	\left\|
	\left[
	\mathbf W_{i,:},\,
	[\mathbf v]_{i}
	\right]
	\right\|_2>0
	\right\}.
	\label{eq:active_support_induced}
\end{equation}

Accordingly, the final selection matrix is obtained as
$\mathbf P_{\mathcal S^\star}
=
[\mathbf e_{s_1},\ldots,\mathbf e_{s_{|\mathcal S^\star|}}]$,
where $\mathcal S^\star=\{s_1,\ldots,s_{|\mathcal S^\star|}\}$.
After the pruning--refitting procedure terminates, the support satisfies $|\mathcal S^\star|=N_{\mathrm t}$, and the corresponding reduced-dimensional BF and AN variables can be written as
$\mathbf W=\mathbf P_{\mathcal S^\star}\widetilde{\mathbf W}_{\mathcal S^\star}$ and
$\mathbf v=\mathbf P_{\mathcal S^\star}\widetilde{\mathbf v}_{\mathcal S^\star}$.

Problem~\eqref{eq:sr_sparse_problem_brief} is a mixed discrete--continuous non-convex optimization problem.
The difficulty arises from the non-concave SR objective, the joint coupling between the BF matrix and AN vector, and the row-sparsity constraint in \eqref{eq:sr_sparse_problem_brief:c}, which implicitly determines the active FA-port support among the $L$ candidate ports.
An exhaustive search over all possible supports is computationally prohibitive for large-scale FAS.
Therefore, a tractable and hardware-consistent solution to Problem~\eqref{eq:sr_sparse_problem_brief} will be developed in Section~\ref{section_BF_AN_optimization}, where the continuous BF/AN design and the discrete FA-port selection are handled through an AO-based prune--refit framework.

\section{Proposed FA-Assisted Secure Transmission Design}
\label{section_BF_AN_optimization}
This section develops the proposed FA-assisted secure transmission design by following the algorithmic flow induced by Problem~\eqref{eq:sr_sparse_problem_brief}. We first derive the fully digital BF--AN solution for a fixed active-port set, then perform power allocation and stream balancing, next update the support through progressive pruning, and finally realize the balanced digital target by a hybrid RF--baseband architecture over the selected ports. The overall procedure and its asymptotic complexity are summarized at the end of this section.

\subsection{Fully Digital BF--AN Design with a Fixed Active-Port Set}

Given an active-port set $\mathcal S$ with selection matrix $\mathbf P_{\mathcal S}$ that extracts the selected FA ports from the $L$ candidate ports of the discretized FAS aperture, define the noise-whitened equivalent channels $\bar{\mathbf H}_{\mathrm U,\mathcal S}=\bar{\mathbf H}_{\mathrm U}\mathbf P_{\mathcal S}\in\mathbb C^{N_{\mathrm u}\times|\mathcal S|}$ and $\bar{\mathbf H}_{\mathrm E,\mathcal S}=\bar{\mathbf H}_{\mathrm E}\mathbf P_{\mathcal S}\in\mathbb C^{N_{\mathrm e}\times|\mathcal S|}$.
Let $\tilde{\mathbf W}_{\mathcal S}\in\mathbb C^{|\mathcal S|\times K}$ and $\tilde{\mathbf v}_{\mathcal S}\in\mathbb C^{|\mathcal S|\times 1}$ denote the fully digital dense BF matrix and AN vector on the $|\mathcal S|$ active ports. Using a BCD procedure, the SR
\(R_s(\widetilde{\mathbf W}_{\mathcal S},\widetilde{\mathbf v}_{\mathcal S})\)
is therefore reformulated as
\begin{equation}
	\begin{aligned}
		 & {R}_\text{s}({\tilde{\mathbf W}_{\mathcal S}},{\tilde{\mathbf{v}}}_{\mathcal{S}},\bar{\mathbf H}_{\mathrm U,\mathsf{\mathcal S}},\bar{\mathbf{H}}_{\text{E},\mathcal{S}}) =                                                                           \\
		 & \underbrace{\log \det \left(
			{\mathbf{I}}_{N_{\text{u}}}+\frac{
				{\bar{\mathbf H}_{\mathrm U,\mathsf{\mathcal S}}} {\tilde{\mathbf W}_{\mathcal S}} {\tilde{\mathbf W}_{\mathcal S}}^{\text{H}} {\bar{\mathbf H}_{\mathrm U,\mathsf{\mathcal S}}}^{\text{H}}
			}{
				{\mathbf{I}}_{N_{\text{u}}} + {\bar{\mathbf H}_{\mathrm U,\mathsf{\mathcal S}}} {\tilde{\mathbf{v}}}_{\mathcal{S}} {\tilde{\mathbf{v}}}_{\mathcal{S}}^{\text{H}} {\bar{\mathbf H}_{\mathrm U,\mathsf{\mathcal S}}}^{\text{H}}
			}
		\right)}_{r_1}                                                                                                                                                                                                                                           \\
		 & + \underbrace{\log \det \left({\mathbf{I}}_{N_{\text{e}}} + \bar{\mathbf{H}}_{\text{E},\mathcal{S}} {\tilde{\mathbf{v}}}_{\mathcal{S}} {\tilde{\mathbf{v}}}_{\mathcal{S}}^{\text{H}} \bar{\mathbf{H}}_{\text{E},\mathcal{S}}^{\text{H}}\right)}_{r_2} \\
		 & - \underbrace{\log \det \left({\mathbf{I}}_{N_{\text{e}}} + \bar{\mathbf{H}}_{\text{E},\mathcal{S}}\, \mathbf{T}_{\mathcal{S}}\mathbf{T}_{\mathcal{S}}^{\text{H}}\, \bar{\mathbf{H}}_{\text{E},\mathcal{S}}^{\text{H}}\right)}_{r_3}
	\end{aligned}
	\label{eq:secrecy_rate_reformulated}
\end{equation}
Here, \(r_1\) denotes Bob's effective information rate in the presence of AN, while \(r_3-r_2\) is Eve's effective information rate; hence the raw SR is expressed as \(r_1+r_2-r_3\). $\mathbf{T}_{\mathcal{S}}\in\mathbb{C}^{|\mathsf{\mathcal S}|\times (K+1)}$ is defined as
\begin{equation}
	\label{eq:T_def}
	\mathbf{T}_{\mathcal{S}}
	\triangleq \big[\,\tilde{\mathbf{W}}_{\mathcal{S}},\tilde{\mathbf{v}}_{\mathcal{S}}\,\big],
\end{equation}
with $
	\mathbf{T}_{\mathcal{S}}\mathbf{T}_{\mathcal{S}}^{\text{H}}
	= \tilde{\mathbf{W}}_{\mathcal{S}}\tilde{\mathbf{W}}_{\mathcal{S}}^{\text{H}}
	+ \tilde{\mathbf{v}}_{\mathcal{S}}\tilde{\mathbf{v}}_{\mathcal{S}}^{\text{H}}.
$

According to \cite[Lemma~4.1]{shi2015Secure} and the Fenchel conjugate construction in \cite[Example~11.7]{boyd2004convex}, the variational form of the Bob-side log-determinant term $r_1$ is
\begin{equation}
	\label{r1_variational}
	{{r}_{1}}
	=\underset{{\mathbf Q}_{\text{\text{B}}}\succ 0,\,{\mathbf J}_{\text{\text{B}}}}{\max}\,
	\log\det({\mathbf Q}_{\text{\text{B}}})
	-\operatorname{Tr}\!\big({\mathbf Q}_{\text{\text{B}}}\,{\mathbf G}_{\text{\text{B}}}(\mathbf U_{\text{\text{\text{B}}}},\tilde{\mathbf W}_{\mathcal S},\tilde{\mathbf v}_{\mathcal S})\big)
	+K,
\end{equation}
where ${\mathbf Q}_{\text B}\succ 0$ and ${\mathbf J}_{\text B}$ are auxiliary variables, and ${\mathbf G}_{\text B}(\cdot)$ denotes the mean-square error (MSE) matrix to be specified below.

At Bob, the interference-plus-noise covariance is defined as
\begin{equation}
	\label{eq:Jb_def}
	\mathbf{J}_{\text{\text{B}}}\triangleq \mathbf{I}_{N_{\text{u}}}
	+\bar{\mathbf{H}}_{\text{\text{\text{B}}},\mathcal{S}}\,
	\tilde{\mathbf{v}}_{\mathcal{S}}\tilde{\mathbf{v}}_{\mathcal{S}}^{\text{H}}\,
	\bar{\mathbf{H}}_{\text{\text{\text{B}}},\mathcal{S}}^{\text{H}} .
\end{equation}
For fixed $(\tilde{\mathbf W}_{\mathcal S},\tilde{\mathbf v}_{\mathcal S})$ and $\mathbf J_{\text B}$, the MSE matrix associated with a linear receive filter $\mathbf U_{\text B}\in\mathbb C^{N_{\text u}\times K}$ is
\begin{equation}
	\label{eq:Gb_def}
	\begin{aligned}
		 & \mathbf{G}_{\text B,\mathcal S}(\mathbf{U}_{\text B},\tilde{\mathbf W}_{\mathcal S},\tilde{\mathbf v}_{\mathcal S})        \\
		 & =\big(\mathbf I_{K}-\mathbf U_{\text B}^{\mathrm H}\bar{\mathbf H}_{\text B,\mathcal S}\tilde{\mathbf W}_{\mathcal S}\big)
		\big(\mathbf I_{K}-\mathbf U_{\text B}^{\mathrm H}\bar{\mathbf H}_{\text B,\mathcal S}\tilde{\mathbf W}_{\mathcal S}\big)^{\!\mathrm H}
		+\mathbf U_{\text B}^{\mathrm H}\mathbf J_{\text B}\mathbf U_{\text B}.
	\end{aligned}
\end{equation}

Since $\mathbf J_{\text B}\succ \mathbf 0$, the objective $\operatorname{Tr}\big(\mathbf{G}_{\text B,\mathcal S}\big)$ is strictly convex in $\mathbf U_{\text B}$, and the unique minimizer is
\begin{equation}
	\label{eq:Ub_opt}
	\mathbf U_{\text B}^{\star}
	=\big(\mathbf J_{\text B}+\bar{\mathbf H}_{\text B,\mathcal S}\tilde{\mathbf W}_{\mathcal S}\tilde{\mathbf W}_{\mathcal S}^{\mathrm H}
	\bar{\mathbf H}_{\text B,\mathcal S}^{\mathrm H}\big)^{-1}\bar{\mathbf H}_{\text B,\mathcal S}\tilde{\mathbf W}_{\mathcal S}.
\end{equation}
Substituting \eqref{eq:Ub_opt} into \eqref{eq:Gb_def} yields the minimal MSE (MMSE) matrix
\begin{equation}
	\label{eq:Gb_star}
	\begin{aligned}
		\mathbf G_{\text B,\mathcal S}^{\star}
		 & \triangleq \mathbf{G}_{\text B,\mathcal S}\big(\mathbf U_{\text B}^{\star},\tilde{\mathbf W}_{\mathcal S},\tilde{\mathbf v}_{\mathcal S}\big) \\
		 & =\Big(\mathbf I_{K}+\tilde{\mathbf W}_{\mathcal S}^{\mathrm H}\bar{\mathbf H}_{\text B,\mathcal S}^{\mathrm H}
		\mathbf J_{\text B}^{-1}\bar{\mathbf H}_{\text B,\mathcal S}\tilde{\mathbf W}_{\mathcal S}\Big)^{-1}.
	\end{aligned}
\end{equation}
This sequence of definitions fixes ${\mathbf J}_{\text B}$, specifies ${\mathbf G}_{\text B, \mathcal S}(\cdot)$, and identifies the optimal receive filter $\mathbf U_{\text B}^{\star}$, which together instantiate the variational representation in \eqref{r1_variational}.

Invoking the concave–convex conjugate identity of the log-determinant, for any $\mathbf G\succ \mathbf 0$,
$
	-\log\det(\mathbf G)
	=\max_{\mathbf Q\succ \mathbf 0}\ \log\det(\mathbf Q)-\operatorname{Tr}(\mathbf Q\,\mathbf G)+K.
$
Since the objective is strictly concave in $\mathbf Q$, the maximizer is unique and satisfies $\mathbf Q^{\star}=\mathbf G^{-1}$.
Substituting $\mathbf G=\mathbf G_{\text B,\mathcal S}^{\star}$ gives the optimizer associated with \eqref{r1_variational}:
\begin{equation}
	\label{eq:Qb_star}
	\mathbf{Q}_{\text{B}}^{\star}=\big(\mathbf{G}_{\text{B},\mathcal S}^{\star}\big)^{-1}.
\end{equation}

Similarly, the Eve AN-only covariance log-det term $r_2$ admits the scalar variational representation
\begin{equation}
	\label{eq:r2_variational_new}
	r_2
	=\max_{\mathbf u_{\mathrm E},\,Q_{\mathrm E,\mathcal S}>0}
	\ \log Q_{\mathrm E,\mathcal S}
	- Q_{\mathrm E,\mathcal S}\,G_{\mathrm E,\mathcal S}
	+ 1.
\end{equation}
Here the auxiliary variables are the linear receive filter
\begin{equation}
	\label{eq:Ue_def_new}
	\mathbf u_{\mathrm E}
	=\big(\mathbf I_{N_{\mathrm e}}
	+\bar{\mathbf H}_{\mathrm E,\mathcal S}\,\tilde{\mathbf v}_{\mathcal S}\tilde{\mathbf v}_{\mathcal S}^{\mathrm H}\bar{\mathbf H}_{\mathrm E,\mathcal S}^{\mathrm H}\big)^{-1}
	\bar{\mathbf H}_{\mathrm E,\mathcal S}\tilde{\mathbf v}_{\mathcal S}\in\mathbb C^{N_{\mathrm e}\times 1},
\end{equation}
and the associated MSE
\begin{equation}
	\label{eq:Ge_def_new}
	G_{\mathrm E,\mathcal S}
	=\big|\,1-\mathbf u_{\mathrm E}^{\mathrm H}\bar{\mathbf H}_{\mathrm E,\mathcal S}\tilde{\mathbf v}_{\mathcal S}\,\big|^{2}
	+\|\mathbf u_{\mathrm E}\|_{2}^{2}.
\end{equation}
Maximization of \eqref{eq:r2_variational_new} with respect to the scalar weight yields
\begin{equation}
	\label{eq:Qe_star_new}
	Q_{\mathrm E,\mathcal S}^{\star}=\big(G_{\mathrm E,\mathcal S}\big)^{-1},
\end{equation}
and, for fixed $\tilde{\mathbf v}_{\mathcal S}$ and $\bar{\mathbf H}_{\mathrm E,\mathcal S}$, the maximizing filter is
\begin{equation}
	\label{eq:ue_star_new}
	\mathbf u_{\mathrm E}^{\star}
	=\big(\mathbf I_{N_{\mathrm e}}
	+\bar{\mathbf H}_{\mathrm E,\mathcal S}\,\tilde{\mathbf v}_{\mathcal S}\tilde{\mathbf v}_{\mathcal S}^{\mathrm H}\bar{\mathbf H}_{\mathrm E,\mathcal S}^{\mathrm H}\big)^{-1}
	\bar{\mathbf H}_{\mathrm E,\mathcal S}\tilde{\mathbf v}_{\mathcal S}.
\end{equation}
Substituting \eqref{eq:Qe_star_new}–\eqref{eq:ue_star_new} into \eqref{eq:r2_variational_new} reproduces the original Eve-side log term and, for fixed auxiliaries, yields a quadratic surrogate in $\tilde{\mathbf v}_{\mathcal S}$.

Next, for the Eve signal-plus-AN covariance log-det term $r_3$, observe that
$
	r_3=\log\det\!\Big(\mathbf I_{N_{\mathrm e}}
	+\bar{\mathbf H}_{\mathrm E,\mathcal S}\,\mathbf T_{\mathcal S}\mathbf T_{\mathcal S}^{\mathrm H}\bar{\mathbf H}_{\mathrm E,\mathcal S}^{\mathrm H}\Big)$,
and introduce
\begin{equation}
	\label{eq:GZ_def_here}
	\mathbf G_{\mathrm Z,\mathcal S}
	\triangleq
	\mathbf I_{N_{\mathrm e}}
	+\bar{\mathbf H}_{\mathrm E,\mathcal S}\,\mathbf T_{\mathcal S}\mathbf T_{\mathcal S}^{\mathrm H}\bar{\mathbf H}_{\mathrm E,\mathcal S}^{\mathrm H}.
\end{equation}
By the concave–convex conjugate identity for the log–determinant, $-\log\det(\mathbf G)=\max_{\mathbf Q\succ \mathbf 0}\,\log\det(\mathbf Q)-\operatorname{Tr}(\mathbf Q\,\mathbf G)+\dim(\mathbf G)$, $-r_3$ admits the matrix variational representation
\begin{equation}
	\label{eq:-r3_variational_final}
	-\,r_3
	=\max_{\mathbf Q_{\mathrm Z,\mathcal S}\succ \mathbf 0}\;
	\log\det(\mathbf Q_{\mathrm Z,\mathcal S})
	-\operatorname{Tr}\!\big(\mathbf Q_{\mathrm Z,\mathcal S}\,\mathbf G_{\mathrm Z,\mathcal S}\big)
	+ N_{\mathrm e},
\end{equation}
whose unique maximizer is
\begin{equation}
	\label{eq:QZ_star_final}
	\mathbf Q_{\mathrm Z,\mathcal S}^{\star}=\mathbf G_{\mathrm Z,\mathcal S}^{-1}.
\end{equation}
With the auxiliaries $\{\mathbf U_{\text B},\mathbf Q_{\text B}\}$, $\{\mathbf u_{\mathrm E},Q_{\mathrm E,\mathcal S}\}$, and $\mathbf Q_{\mathrm Z,\mathcal S}$ fixed, the  $r_1$,  $r_2$, and $r_3$ admit quadratic surrogates in $(\tilde{\mathbf W}_{\mathcal S},\tilde{\mathbf v}_{\mathcal S})$ that separate design variables from constants as

\begin{equation}
	\label{eq:r1_rewrite_sep}
	\begin{aligned}
		r_1
		 & = -\,\operatorname{Tr}\!\big(\tilde{\mathbf W}_{\mathcal S}^{\mathrm H}\mathbf F_{b,\mathcal S}\tilde{\mathbf W}_{\mathcal S}\big)
		+ 2\,\Re\!\big\{\operatorname{Tr}(\mathbf R_{w,\mathcal S}^{\mathrm H}\tilde{\mathbf W}_{\mathcal S})\big\}                           \\
		 & \quad +\underbrace{\log\det(\mathbf Q_{\text B}) + K
			- \operatorname{Tr}(\mathbf Q_{\text B})
			- \operatorname{Tr}\!\big(\mathbf Q_{\text B}\,\mathbf U_{\text B}^{\mathrm H}\mathbf J_{\text B}\mathbf U_{\text B}\big)}_{c_{\text B}} .
	\end{aligned}
\end{equation}

\begin{equation}
	\label{eq:r2_rewrite_sep}
	\begin{aligned}
		r_2
		 & = -\,\tilde{\mathbf v}_{\mathcal S}^{\mathrm H}\mathbf F_{e,\mathcal S}\tilde{\mathbf v}_{\mathcal S}
		+ 2\,\Re\!\big\{\mathbf r_{v,\mathcal S}^{\mathrm H}\tilde{\mathbf v}_{\mathcal S}\big\}                 \\
		 & \quad + \underbrace{\log Q_{\mathrm E,\mathcal S} + 1
			- Q_{\mathrm E,\mathcal S}\!\left(1+\|\mathbf u_{\mathrm E}\|_2^2\right)}_{c_{\mathrm E}} .
	\end{aligned}
\end{equation}

\begin{equation}
	\label{eq:r3_rewrite_sep}
	\begin{aligned}
		r_3
		 & = -\,\operatorname{Tr}\!\big(\tilde{\mathbf W}_{\mathcal S}^{\mathrm H}\mathbf C_{\mathcal S}\tilde{\mathbf W}_{\mathcal S}\big)
		- \tilde{\mathbf v}_{\mathcal S}^{\mathrm H}\mathbf C_{\mathcal S}\tilde{\mathbf v}_{\mathcal S}                                    \\
		 & \quad + \underbrace{N_{\mathrm e} + \log\det(\mathbf Q_{\mathrm Z,\mathcal S})
			- \operatorname{Tr}(\mathbf Q_{\mathrm Z,\mathcal S})}_{c_{\mathrm Z}}.
	\end{aligned}
\end{equation}

The Bob-side curvature and linear term are defined as
\begin{equation}
	\label{eq:Fb_def}
	\mathbf F_{b,\mathcal S}
	\triangleq
	\bar{\mathbf H}_{\text B,\mathcal S}^{\mathrm H}\,\mathbf U_{\text B}\,\mathbf Q_{\text B}\,\mathbf U_{\text B}^{\mathrm H}\,\bar{\mathbf H}_{\text B,\mathcal S}
	\succeq \mathbf 0,
\end{equation}
\begin{equation}
	\label{eq:Rw_def}
	\mathbf R_{w,\mathcal S}
	\triangleq
	\bar{\mathbf H}_{\text B,\mathcal S}^{\mathrm H}\,\mathbf U_{\text B}\,\mathbf Q_{\text B},
\end{equation}
the Eve-side counterparts are
\begin{equation}
	\label{eq:Fe_def}
	\mathbf F_{e,\mathcal S}
	\triangleq
	\bar{\mathbf H}_{\text E,\mathcal S}^{\mathrm H}\,\mathbf u_{\mathrm E}\,Q_{\mathrm E,\mathcal S}\,\mathbf u_{\mathrm E}^{\mathrm H}\,\bar{\mathbf H}_{\text E,\mathcal S}
	\succeq \mathbf 0,
\end{equation}
\begin{equation}
	\label{eq:rv_def}
	\mathbf r_{v,\mathcal S}
	\triangleq
	\bar{\mathbf H}_{\text E,\mathcal S}^{\mathrm H}\,\mathbf u_{\mathrm E}\,Q_{\mathrm E,\mathcal S},
\end{equation}
and the aggregate Eve-side curvature is
\begin{equation}
	\label{eq:C_def}
	\mathbf C_{\mathcal S}
	\triangleq
	\bar{\mathbf H}_{\text E,\mathcal S}^{\mathrm H}\,\mathbf Q_{\mathrm Z,\mathcal S}\,\bar{\mathbf H}_{\text E,\mathcal S}
	\succeq \mathbf 0.
\end{equation}


Dropping the constants \(c_{\text B},c_{\mathrm E},c_{\mathrm Z}\), the SR maximization over the design variables reduces to the convex quadratic program (CQP)
\begin{subequations}
	\label{eq:quad_prob_ieee_corrected}
	\begin{align}
		\min_{\tilde{\mathbf{W}}_{\mathcal S},\,\tilde{\mathbf{v}}_{\mathcal S}}\ \
		 & \operatorname{Tr}\!\big(\tilde{\mathbf{W}}_{\mathcal S}^{\mathrm H}\mathbf{A}_{\mathcal S}\tilde{\mathbf{W}}_{\mathcal S}\big)
		-2\,\Re\!\big\{\operatorname{Tr}(\mathbf{R}_{w,\mathcal S}^{\mathrm H}\tilde{\mathbf{W}}_{\mathcal S})\big\} \notag               \\
		 & \quad + \tilde{\mathbf{v}}_{\mathcal S}^{\mathrm H}\mathbf{B}_{\mathcal S}\tilde{\mathbf{v}}_{\mathcal S}
		-2\,\Re\!\big\{\mathbf{r}_{v,\mathcal S}^{\mathrm H}\tilde{\mathbf{v}}_{\mathcal S}\big\},
		\label{eq:quad_prob_ieee_corrected:a}                                                                                             \\
		\text{s.t. }\
		 & \operatorname{Tr}(\mathbf{T}_{\mathcal S}\mathbf{T}_{\mathcal S}^{\mathrm H}) \le P_{\mathrm t}.
		\label{eq:quad_prob_ieee_corrected:b}
	\end{align}
\end{subequations}
with
\begin{equation}
	\label{eq:A_B_defs_here}
	\mathbf A_{\mathcal S}\ \triangleq\ \mathbf F_{b,\mathcal S}+\mathbf C_{\mathcal S},
	\qquad
	\mathbf B_{\mathcal S}\ \triangleq\ \mathbf F_{b,\mathcal S}+\mathbf F_{e,\mathcal S}+\mathbf C_{\mathcal S}.
\end{equation}

Introducing a Lagrange multiplier \(\lambda\ge 0\) for the total–power constraint \eqref{eq:quad_prob_ieee_corrected:b}, the Lagrangian is
\begin{equation}
	\label{eq:L_ieee}
	\begin{aligned}
		\mathcal{L}(\tilde{\mathbf{W}}_{\mathcal S},\tilde{\mathbf{v}}_{\mathcal S};\lambda)=\,
		 & \operatorname{Tr}(\tilde{\mathbf{W}}_{\mathcal S}^{\mathrm H}\mathbf{A}_{\mathcal S}\tilde{\mathbf{W}}_{\mathcal S})
		-2\,\Re\!\big\{\operatorname{Tr}(\mathbf{R}_{w,\mathcal S}^{\mathrm H}\tilde{\mathbf{W}}_{\mathcal S})\big\}            \\
		 & + \tilde{\mathbf{v}}_{\mathcal S}^{\mathrm H}\mathbf{B}_{\mathcal S}\tilde{\mathbf{v}}_{\mathcal S}
		-2\,\Re\!\big\{\mathbf{r}_{v,\mathcal S}^{\mathrm H}\tilde{\mathbf{v}}_{\mathcal S}\big\}                               \\
		 & + \lambda\!\left(\operatorname{Tr}(\mathbf{T}_{\mathcal S}\mathbf{T}_{\mathcal S}^{\mathrm H})-P_{\mathrm t}\right).
	\end{aligned}
\end{equation}
Minimizing \eqref{eq:L_ieee} with respect to \(\tilde{\mathbf{W}}_{\mathcal S}\) and \(\tilde{\mathbf{v}}_{\mathcal S}\) gives
\begin{equation}
	\label{eq:normal_W}
	(\mathbf{A}_{\mathcal S}+\lambda\mathbf{I}_{|\mathcal S|})\,\tilde{\mathbf{W}}_{\mathcal S}
	=\mathbf{R}_{w,\mathcal S},
\end{equation}
\begin{equation}
	\label{eq:normal_v}
	(\mathbf{B}_{\mathcal S}+\lambda\mathbf{I}_{|\mathcal S|})\,\tilde{\mathbf{v}}_{\mathcal S}
	=\mathbf{r}_{v,\mathcal S}.
\end{equation}

Since \(\mathbf{A}_{\mathcal S}\succeq\mathbf 0\), \(\mathbf{B}_{\mathcal S}\succeq\mathbf 0\), the matrices
\(\mathbf{A}_{\mathcal S}+\lambda\mathbf{I}_{|\mathcal S|}\) and \(\mathbf{B}_{\mathcal S}+\lambda\mathbf{I}_{|\mathcal S|}\) are positive definite for any \(\lambda>0\), and the unique minimizers are
\begin{equation}
	\label{eq:closed_forms_matrix_final}
	\tilde{\mathbf{W}}_{\mathcal S}(\lambda)
	=(\mathbf{A}_{\mathcal S}+\lambda\mathbf{I}_{|\mathcal S|})^{-1}\mathbf{R}_{w,\mathcal S},
\end{equation}
\begin{equation}
	\label{eq:closed_forms_vector_final}
	\tilde{\mathbf{v}}_{\mathcal S}(\lambda)
	=(\mathbf{B}_{\mathcal S}+\lambda\mathbf{I}_{|\mathcal S|})^{-1}\mathbf{r}_{v,\mathcal S}.
\end{equation}
Equivalently,
\begin{equation}
	\label{eq:inner_min}
	(\tilde{\mathbf{W}}_{\mathcal S}(\lambda),\tilde{\mathbf{v}}_{\mathcal S}(\lambda))
	\in\arg\min_{\tilde{\mathbf{W}}_{\mathcal S},\,\tilde{\mathbf{v}}_{\mathcal S}}
	\ \mathcal{L}(\tilde{\mathbf{W}}_{\mathcal S},\tilde{\mathbf{v}}_{\mathcal S};\lambda).
\end{equation}

The dual function is defined by the inner minimum
\begin{equation}
	\label{eq:dual_fun}
	g(\lambda)\triangleq \min_{\tilde{\mathbf{W}}_{\mathcal S},\,\tilde{\mathbf{v}}_{\mathcal S}}
	\ \mathcal{L}(\tilde{\mathbf{W}}_{\mathcal S},\tilde{\mathbf{v}}_{\mathcal S};\lambda),
\end{equation}
and evaluates in closed form as
\begin{equation}
	\label{eq:dual_fun_value}
	\begin{aligned}
		g(\lambda)=\,
		 & -\left\|(\mathbf{A}_{\mathcal S}+\lambda\mathbf{I}_{|\mathcal S|})^{-1/2}\mathbf{R}_{w,\mathcal S}\right\|_{\operatorname{F}}^{2} \\
		 & -\left\|(\mathbf{B}_{\mathcal S}+\lambda\mathbf{I}_{|\mathcal S|})^{-1/2}\mathbf{r}_{v,\mathcal S}\right\|_{2}^{2}
		-\lambda P_{\mathrm t}.
	\end{aligned}
\end{equation}

The dual function in \eqref{eq:dual_fun} can be maximized by a direct one-dimensional search in \(\lambda\), such as bisection on the complementary-slackness residual. This strategy repeatedly inverts \((\mathbf A_{\mathcal S}+\lambda\mathbf I)\) and \((\mathbf B_{\mathcal S}+\lambda\mathbf I)\) for each trial \(\lambda\). When \(\mathbf A_{\mathcal S}\) or \(\mathbf B_{\mathcal S}\) is ill conditioned or nearly singular, the residual becomes steep near \(\lambda=0\), which degrades numerical stability and amplifies finite-precision effects. The per-iteration cost is also dominated by fresh matrix factorizations with little reuse. These considerations motivate a diagonalized treatment that reveals per-mode monotonicity and enables a robust update of \(\lambda\). The complementary slackness condition is

\begin{equation}
	\label{eq:comp_slackness_trace}
	\lambda\Big(
	\underbrace{\operatorname{Tr}(\tilde{\mathbf{W}}_{\mathcal S}\tilde{\mathbf{W}}_{\mathcal S}^{\mathrm H})
		+\operatorname{Tr}(\tilde{\mathbf v}_{\mathcal S}\tilde{\mathbf v}_{\mathcal S}^{\mathrm H})}%
	_{\operatorname{Tr}(\mathbf{T}_{\mathcal S}\mathbf{T}_{\mathcal S}^{\mathrm H})}
	- P_{\mathrm t}\Big)=0 .
\end{equation}

\subsection{Power Allocation and Stream Balancing}
The dual water level \(\lambda\) is determined in the eigenbases of \(\mathbf A_{\mathcal S}\) and \(\mathbf B_{\mathcal S}\). Diagonalize the Hermitian curvature matrices as
\begin{equation}
	\label{eq:eigsA_gswf}
	\mathbf A_{\mathcal S}=\mathbf Z_{A}\boldsymbol{\Xi}_{A}\mathbf Z_{A}^{\mathrm H},\qquad
	\boldsymbol{\Xi}_{A}=\operatorname{diag}(\xi_{A,1},\ldots,\xi_{A,|\mathcal S|}),
\end{equation}
and
\begin{equation}
	\label{eq:eigsB_gswf}
	\mathbf B_{\mathcal S}=\mathbf Z_{B}\boldsymbol{\Xi}_{B}\mathbf Z_{B}^{\mathrm H},\qquad
	\boldsymbol{\Xi}_{B}=\operatorname{diag}(\xi_{B,1},\ldots,\xi_{B,|\mathcal S|}),
\end{equation}
where \(\mathbf Z_A,\mathbf Z_B\) collect orthonormal eigenvectors and \(\xi_{A,i},\xi_{B,i}\ge 0\) are the eigenvalues. Rotate the linear terms into these eigenbases as
\begin{equation}
	\label{eq:rot_rhs_gswf}
	\widehat{\mathbf R}\triangleq \mathbf Z_{A}^{\mathrm H}\mathbf R_{w,\mathcal S},\qquad
	\widehat{\mathbf r}\triangleq \mathbf Z_{B}^{\mathrm H}\mathbf r_{v,\mathcal S}.
\end{equation}

With \eqref{eq:eigsA_gswf}–\eqref{eq:rot_rhs_gswf}, the closed-form minimizers in \eqref{eq:closed_forms_matrix_final}–\eqref{eq:closed_forms_vector_final} take mode-wise shrinkage forms,
\begin{equation}
	\label{eq:W_spec_gswf}
	\tilde{\mathbf W}_{\mathcal S}(\lambda)
	=\mathbf Z_{A}\,\operatorname{diag}\!\big((\xi_{A,i}+\lambda)^{-1}\big)\,\widehat{\mathbf R},
\end{equation}
and
\begin{equation}
	\label{eq:v_spec_gswf}
	\tilde{\mathbf v}_{\mathcal S}(\lambda)
	=\mathbf Z_{B}\,\operatorname{diag}\!\big((\xi_{B,i}+\lambda)^{-1}\big)\,\widehat{\mathbf r}.
\end{equation}
The corresponding powers separate across spectral modes:
\begin{equation}
	\label{eq:power_W_gswf}
	\big\|\tilde{\mathbf W}_{\mathcal S}(\lambda)\big\|_{\operatorname{F}}^{2}
	=\sum_{i=1}^{|\mathcal S|}\frac{\big\|\widehat{\mathbf R}_{i,:}\big\|_{2}^{2}}{(\xi_{A,i}+\lambda)^{2}},
\end{equation}
and
\begin{equation}
	\label{eq:power_v_gswf}
	\big\|\tilde{\mathbf v}_{\mathcal S}(\lambda)\big\|_{2}^{2}
	=\sum_{i=1}^{|\mathcal S|}\frac{|\widehat r_{i}|^{2}}{(\xi_{B,i}+\lambda)^{2}}.
\end{equation}

Enforcing the total-power constraint reduces to solving the scalar equation

\begin{equation}
	\label{eq:waterfill_eq_gswf}
	\sum_{i=1}^{|\mathcal S|}
	\frac{\big\|\widehat{\mathbf R}_{i,:}\big\|_{2}^{2}}{(\xi_{A,i}+\lambda)^{2}}
	\;+\;
	\sum_{i=1}^{|\mathcal S|}
	\frac{|\widehat r_{i}|^{2}}{(\xi_{B,i}+\lambda)^{2}}
	\;=\; P_{\mathrm t}.
\end{equation}

The left-hand side of \eqref{eq:waterfill_eq_gswf} is continuous and strictly decreasing on \([0,\infty)\), so a unique solution \(\lambda^\star\) exists. A safeguarded bisection directly applies to \eqref{eq:waterfill_eq_gswf}; after the one-time eigendecompositions in \eqref{eq:eigsA_gswf}–\eqref{eq:eigsB_gswf}, each iteration evaluates only the two spectral sums and requires no additional matrix inversions. Working in the diagonalized basis also improves numerical stability when \(\mathbf A_{\mathcal S}\) or \(\mathbf B_{\mathcal S}\) is ill-conditioned.

Substituting \(\lambda^\star\) into \eqref{eq:W_spec_gswf}–\eqref{eq:v_spec_gswf} gives the primal updates
\begin{equation}
	\label{eq:W_star_gswf}
	\tilde{\mathbf W}_{\mathcal S}^\star=\tilde{\mathbf W}_{\mathcal S}(\lambda^\star),
	\quad
	\tilde{\mathbf v}_{\mathcal S}^\star=\tilde{\mathbf v}_{\mathcal S}(\lambda^\star),
\end{equation}
which meet the power constraint with equality and minimize the quadratic surrogate for the fixed auxiliaries.

Although \eqref{eq:W_star_gswf} minimizes the surrogate, it may concentrate data power on a few streams when \(K>1\). 
A right-unitary post-processing is therefore applied to balance the per-stream powers without altering the covariance or the total power. 
Let \(\mathbf S_{\mathrm d}\triangleq\widetilde{\mathbf W}_{\mathcal S}^{\star}(\widetilde{\mathbf W}_{\mathcal S}^{\star})^{\mathrm H}\) with total power \(P_{\mathrm d}\triangleq\operatorname{Tr}(\mathbf S_{\mathrm d})\). 
For any unitary \(\boldsymbol{\Omega}\in\mathbb C^{K\times K}\), the rotated precoder \(\widehat{\mathbf W}_{\mathcal S}=\widetilde{\mathbf W}_{\mathcal S}^{\star}\boldsymbol{\Omega}\) satisfies \(\widehat{\mathbf W}_{\mathcal S}\widehat{\mathbf W}_{\mathcal S}^{\mathrm H}=\mathbf S_{\mathrm d}\) and \(\operatorname{Tr}(\widehat{\mathbf W}_{\mathcal S}\widehat{\mathbf W}_{\mathcal S}^{\mathrm H})=P_{\mathrm d}\), hence the secrecy objective remains unchanged. 
In this work, \(\boldsymbol{\Omega}\) is chosen as the normalized \(K\)-point DFT matrix \(\boldsymbol{\Omega}_{\mathrm{bal}}\), which provides a deterministic low-complexity rotation for mitigating stream-power imbalance.

The final balanced update keeps the AN vector unchanged and rotates only the data precoder:
\begin{equation}
	\label{eq:balanced_updates}
	\tilde{\mathbf W}_{\mathcal S}^{\mathrm{bal}}
	=\tilde{\mathbf W}_{\mathcal S}^{\star}\boldsymbol{\Omega}_{\mathrm{bal}},
	\qquad
	\tilde{\mathbf v}_{\mathcal S}^{\mathrm{bal}}=\tilde{\mathbf v}_{\mathcal S}^{\star}.
\end{equation}
This post-processing exactly preserves \(\mathbf S_{\mathrm d}\) and the total power, improves the conditioning at Bob’s linear receiver, and facilitates the subsequent hybrid realization.

\subsection{Active-Port Selection via Progressive Pruning}
\label{sec:fa_port_selection}

After obtaining the fully digital BF--AN update on a fixed support and applying the balancing step, the active-port set is refined in the digital domain through a progressive prune--refit procedure. At each stage, the BF/AN variables are first refitted on the current support, then row-energy scores are evaluated, and finally the weakest ports are removed; no intermediate HBF fitting is involved during this staged pruning process.

This subsection develops a selection rule for FA ports based on the quadratic surrogate in
\eqref{eq:quad_prob_ieee_corrected} with Lagrangian \eqref{eq:L_ieee}, written in terms of
$\mathbf T_{\mathcal S}$ in \eqref{eq:T_def} for a given support
$\mathcal S\subseteq\{1,\ldots,L\}$.

For some multiplier $\lambda\ge 0$, the stationarity conditions
\eqref{eq:normal_W}–\eqref{eq:normal_v} can be expressed in gradient form as
\begin{equation}
	\label{eq:grad_W_block}
	\nabla_{\tilde{\mathbf W}_{\mathcal S}^{\!*}}\mathcal J
	= \mathbf A_{\mathcal S}\tilde{\mathbf W}_{\mathcal S}-\mathbf R_{w,\mathcal S}
	= -\lambda\,\tilde{\mathbf W}_{\mathcal S},
\end{equation}
\begin{equation}
	\label{eq:grad_v_block}
	\nabla_{\tilde{\mathbf v}_{\mathcal S}^{\!*}}\mathcal J
	= \mathbf B_{\mathcal S}\tilde{\mathbf v}_{\mathcal S}-\mathbf r_{v,\mathcal S}
	= -\lambda\,\tilde{\mathbf v}_{\mathcal S}.
\end{equation}
where $\mathcal J$ denotes the quadratic objective in
\eqref{eq:quad_prob_ieee_corrected:a}.
Let $\mathbf t_{i,:}$ denote the $i$-th row of $\mathbf T_{\mathcal S}$ and define the joint
row-energy score
\begin{equation}
	\label{eq:row_energy_def}
	e_i \;\triangleq\; \|\mathbf t_{i,:}\|_2
	=\big\|\,[\,\tilde{\mathbf W}_{\mathcal S}(i,:),\ \tilde{\mathbf v}_{\mathcal S}(i)\,]\big\|_2.
\end{equation}
From the above gradients, the block-gradient norm associated with row $i$ satisfies
\begin{equation}
	\label{eq:grad_vs_energy}
	\big\|\nabla_i\mathcal J(\mathbf T_{\mathcal S})\big\|_2
	= \lambda\,e_i,
\end{equation}
so ordering rows by $e_i$ is equivalent, up to the common factor $\lambda$, to ordering them
by the corresponding block-gradient norms.

To relate this ranking to a convex sparse model, consider the group-Lasso surrogate
\cite{yuan2006model,jenatton2011structured}
\begin{equation}
	\label{eq:group_lasso_surrogate_final}
	\min_{\tilde{\mathbf W}_{\mathcal S},\,\tilde{\mathbf v}_{\mathcal S}}
	\ \ \mathcal J(\tilde{\mathbf W}_{\mathcal S},\tilde{\mathbf v}_{\mathcal S})
	+\mu\sum_{i=1}^{|\mathcal S|}\|\mathbf t_{i,:}\|_2, \qquad \mu>0,
\end{equation}
with $\mathbf T_{\mathcal S}$ and $\mathbf t_{i,:}$ as above.
Problem \eqref{eq:group_lasso_surrogate_final} is convex and its optimal solutions satisfy
the row-wise KKT conditions \cite[Sec.~5.5]{boyd2004convex}:
\begin{align}
	\label{eq:kkt_active}
	\|\mathbf t_{i,:}^\star\|_2>0 & \Rightarrow
	\nabla_i\mathcal J(\mathbf T_{\mathcal S}^\star)
	= \mu\,\mathbf t_{i,:}^\star/\|\mathbf t_{i,:}^\star\|_2, \\
	\label{eq:kkt_inactive}
	\|\mathbf t_{j,:}^\star\|_2=0 & \Rightarrow
	\|\nabla_j\mathcal J(\mathbf T_{\mathcal S}^\star)\|_2\le \mu.
\end{align}

Combining \eqref{eq:grad_vs_energy} with \eqref{eq:kkt_active}–\eqref{eq:kkt_inactive}
indicates that rows with small $e_i$ are closest to the inactive KKT pattern of
\eqref{eq:group_lasso_surrogate_final}, whereas rows with large $e_i$ are encouraged to remain
active. The scores $e_i$ thus act as importance indicators for the FA ports.

In the proposed selector, the current support $\mathcal S_t$ at stage $t$ is interpreted as
an approximate active set of \eqref{eq:group_lasso_surrogate_final}. Refitting on $\mathcal S_t$
yields a point whose gradients on the retained rows are small, while the rows outside the support
exhibit larger gradients. When such a gradient-norm separation is present, one can associate
$\mathcal S_t$ with an approximate KKT point of \eqref{eq:group_lasso_surrogate_final} for a
suitable $\mu$, so pruning rows with the smallest $e_i$ is consistent with the convex surrogate
and aligns with the behavior of reweighted $\ell_1$ and hard-thresholding strategies
\cite{Candes2008ReweightedL1,Blumensath2009IHT}.

This principle is implemented through a staged prune–refit routine that removes several indices
at each stage.
Let \(\mathcal S_t\) be the current support, \(0<\eta<1\) the pruning ratio, and \(m_{\min}\ge 1\) the minimum batch size. The number of indices removed at stage \(t\) is

\begin{equation}
	\label{eq:batch_size}
	d_t \;\triangleq\; 
	\min\!\Big\{\,|\mathcal S_t|-N_{\mathrm t},\ 
	\max\!\big(m_{\min},\ \lfloor \eta\,(|\mathcal S_t|-N_{\mathrm t}) \rfloor\big)\Big\}.
\end{equation}

Since \(d_t\le |\mathcal S_t|-N_{\mathrm t}\), the updated support satisfies \(|\mathcal S_{t+1}|=|\mathcal S_t|-d_t\ge N_{\mathrm t}\), which prevents over-pruning.
Given the current scores $e_i$, the deletion set is chosen as
\begin{equation}
	\label{eq:delete_set}
	\mathcal D_t \;\in\; \arg\min_{\mathcal D\subseteq\mathcal S_t,\;|\mathcal D|=d_t}\ \sum_{i\in\mathcal D} e_i,
\end{equation}
namely the $d_t$ indices with the smallest row energies, and the support is updated as
\begin{equation}
	\label{eq:support_update}
	\mathcal S_{t+1} \;\leftarrow\; \mathcal S_t \setminus \mathcal D_t.
\end{equation}
Each stage consists of refitting the BF/AN variables on $\mathcal S_t$, computing the scores
$e_i$, and pruning according to \eqref{eq:batch_size}–\eqref{eq:support_update}. The iterations
terminate once $|\mathcal S_t|=N_{\text t}$, followed by a final refit on the terminal support
to reduce shrinkage bias.

The selection matrices follow the support updates. With
$\mathbf P_{\mathcal S}\triangleq\mathbf I_L(:,\mathcal S)$, the initialization is
$\mathcal S_0=\{1,\ldots,L\}$ and $\mathbf P_{\mathcal S_0}=\mathbf I_L$.
After each pruning step in \eqref{eq:support_update}, the selection matrix is updated as
\begin{equation}
	\label{eq:P_update_min}
	\mathbf P_{\mathcal S_{t+1}}
	=\mathbf I_L(:,\mathcal S_{t+1})
	=\mathbf I_L\big(:,\mathcal S_t\setminus\mathcal D_t\big),
\end{equation}
which ensures that subsequent BF/AN updates are always carried out on the currently active
FA ports.

\begin{figure}[!t]
	\centering
	\subfloat[]{%
		\includegraphics[width=1\linewidth]{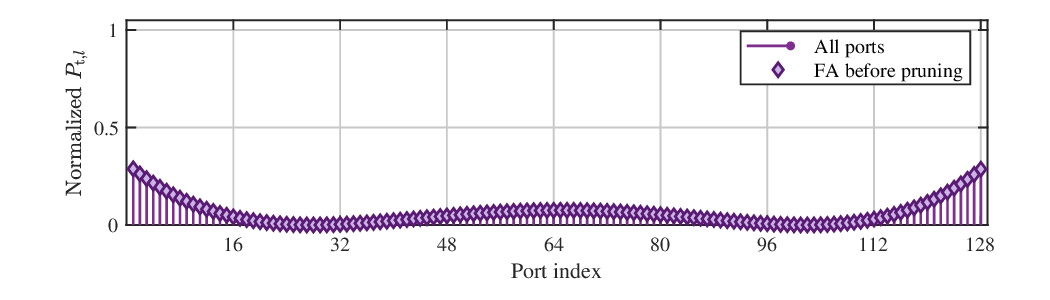}%
		\label{subfig:FA_before_pruning_power}}\\
	\subfloat[]{%
		\includegraphics[width=1\linewidth]{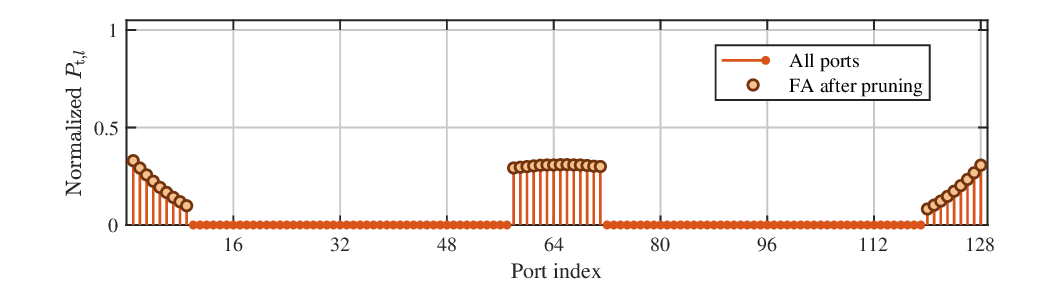}%
		\label{subfig:FA_after_pruning_power}}\\
	\subfloat[]{%
		\includegraphics[width=1\linewidth]{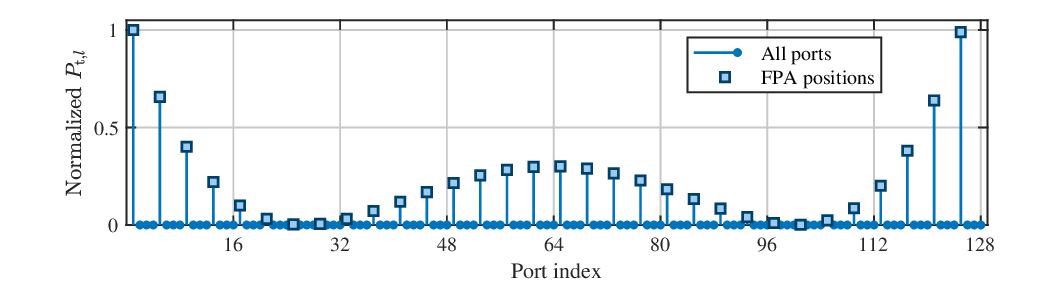}%
		\label{subfig:FPA_array_power}}
		
	\caption{Normalized per-port transmit-power allocation with \(L=128\)
	candidate rail positions and \(N_{\mathrm t}=32\) active ports at
	\(P_{\mathrm t}=10\,\mathrm{dBm}\): (a) FA before pruning; (b) FA after
	pruning; (c) FPA.}
	\label{fig:port_selection_power_allocation}
	
	\vspace{-0.2cm}
\end{figure}

Fig.~\ref{fig:port_selection_power_allocation} further illustrates the physical
effect of the proposed pruning rule by comparing the normalized per-port
transmit-power allocation before pruning, after pruning, and under the FPA
setting. All three cases show larger powers around the aperture center and
two side regions, which agrees with near-field focusing and range-selective
field shaping. However, with fixed FPA positions, some RF-connected antennas
carry very small powers, indicating that the fixed geometry cannot fully place
the available RF-connected ports in high-utility regions. By contrast, under
the same active-port budget \(N_{\mathrm t}\), FA can relocate the activated
ports to more effective positions along the rail. The proposed pruning rule
therefore moves the RF-connected ports toward high-utility regions while
preserving the desired near-field shaping pattern, leading to a more balanced
power distribution among the retained active ports. This is beneficial for
practical implementation since it alleviates the peak-power burden on
individual power amplifiers without increasing the number of active RF chains.

\subsection{Hybrid BF--AN Realization over the Selected Ports}
\label{sec:hbf_realization}

This subsection develops the hybrid RF--baseband realization of the proposed scheme after the active-port selection procedure in Section~\ref{sec:fa_port_selection} returns the final support \(\mathcal S\) with \(|\mathcal S|=N_{\mathrm t}\).
Let \(\widetilde{\mathbf W}_{\mathcal S}^{\mathrm{bal}}\in\mathbb C^{|\mathcal S|\times K}\) denote the balanced fully digital BF target obtained from \eqref{eq:balanced_updates}, and let \(\widetilde{\mathbf v}_{\mathcal S}^{\mathrm{bal}}\in\mathbb C^{|\mathcal S|\times 1}\) denote the corresponding balanced fully digital AN vector.
Following the definition of \(\mathbf T_{\mathcal S}\), the balanced fully digital target is obtained by replacing \(\widetilde{\mathbf W}_{\mathcal S}\) and \(\widetilde{\mathbf v}_{\mathcal S}\) with \(\widetilde{\mathbf W}_{\mathcal S}^{\mathrm{bal}}\) and \(\widetilde{\mathbf v}_{\mathcal S}^{\mathrm{bal}}\), respectively.

We adopt a shared-RF hybrid structure with separate digital BF and AN branches.
The resulting hybrid BF matrix and AN vector are given by
\begin{equation}
	\mathbf W_{\mathrm{HB}}
	\triangleq
	\mathbf F_{\mathrm{RF}}\mathbf F_{\mathrm s},
	\qquad
	\mathbf v_{\mathrm{HB}}
	\triangleq
	\mathbf F_{\mathrm{RF}}\mathbf f_{\mathrm a},
	\label{eq:hbf_bf_an}
\end{equation}
where \(\mathbf W_{\mathrm{HB}}\in\mathbb C^{|\mathcal S|\times K}\), \(\mathbf v_{\mathrm{HB}}\in\mathbb C^{|\mathcal S|\times 1}\), \(\mathbf F_{\mathrm{RF}}\in\mathbb C^{|\mathcal S|\times N_{\mathrm{RF}}}\), \(\mathbf F_{\mathrm s}\in\mathbb C^{N_{\mathrm{RF}}\times K}\), and \(\mathbf f_{\mathrm a}\in\mathbb C^{N_{\mathrm{RF}}\times 1}\).
Since the RF network is implemented by phase shifters, its entries satisfy the constant-modulus constraint
\begin{equation}
	\bigl|[\mathbf F_{\mathrm{RF}}]_{m,n}\bigr|
	=
	\frac{1}{\sqrt{|\mathcal S|}},
	\qquad \forall\,m,n.
	\label{eq:frf_cm}
\end{equation}

To jointly fit the BF and AN targets in the receive domains of Bob and Eve as well as in the transmit domain, we define
\begin{equation}
	\mathbf L_{\mathcal S}
	\triangleq
	\begin{bmatrix}
		\sqrt{w_{\mathrm B}}\,\bar{\mathbf H}_{\mathrm U,\mathcal S} \\
		\sqrt{w_{\mathrm E}}\,\bar{\mathbf H}_{\mathrm E,\mathcal S} \\
		\sqrt{w_{\mathrm T}}\,\mathbf I_{|\mathcal S|}
	\end{bmatrix},
	\label{eq:LS_def}
\end{equation}
where \(w_{\mathrm B},w_{\mathrm E},w_{\mathrm T}>0\) are weighting factors.
The target BF and AN powers are
\begin{equation}
	P_{\mathrm{BF}}^{\star}
	\triangleq
	\operatorname{Tr}\!\left(
	\widetilde{\mathbf W}_{\mathcal S}^{\mathrm{bal}}
	(\widetilde{\mathbf W}_{\mathcal S}^{\mathrm{bal}})^{\mathrm H}
	\right),
	\qquad
	P_{\mathrm{AN}}^{\star}
	\triangleq
	\bigl\|
	\widetilde{\mathbf v}_{\mathcal S}^{\mathrm{bal}}
	\bigr\|_2^2.
	\label{eq:target_powers}
\end{equation}
Defining
\begin{equation}
	\mathbf X
	\triangleq
	\bigl[\mathbf F_{\mathrm s},\,\mathbf f_{\mathrm a}\bigr]
	\in
	\mathbb C^{N_{\mathrm{RF}}\times (K+1)},
	\label{eq:X_def}
\end{equation}
the hybrid realization problem is formulated as
\begin{subequations}\label{eq:hbf_fit_problem}
\begin{align}
	\min_{\mathbf F_{\mathrm{RF}},\,\mathbf F_{\mathrm s},\,\mathbf f_{\mathrm a}}
	\quad
	&
	\left\|
	\mathbf L_{\mathcal S}
	\left(
	[\widetilde{\mathbf W}_{\mathcal S}^{\mathrm{bal}},\,\widetilde{\mathbf v}_{\mathcal S}^{\mathrm{bal}}]
	-
	\mathbf F_{\mathrm{RF}}\mathbf X
	\right)
	\right\|_{\mathrm F}^{2},
	\label{eq:hbf_fit_problem:a}
	\\
	\mathrm{s.t.}\quad
	&
	\bigl|[\mathbf F_{\mathrm{RF}}]_{m,n}\bigr|
	=
	\frac{1}{\sqrt{|\mathcal S|}},
	\qquad \forall\,m,n,
	\label{eq:hbf_fit_problem:b}
	\\
	&
	\operatorname{Tr}\!\left(
	\mathbf W_{\mathrm{HB}}\mathbf W_{\mathrm{HB}}^{\mathrm H}
	\right)
	=
	P_{\mathrm{BF}}^{\star},
	\label{eq:hbf_fit_problem:c}
	\\
	&
	\|\mathbf v_{\mathrm{HB}}\|_2^2
	=
	P_{\mathrm{AN}}^{\star}.
	\label{eq:hbf_fit_problem:d}
\end{align}
\end{subequations}

Problem~\eqref{eq:hbf_fit_problem} is handled by block coordinate descent.
For fixed \(\mathbf F_{\mathrm{RF}}\), define
\begin{equation}
	\mathbf A_{\mathcal S}
	\triangleq
	\mathbf L_{\mathcal S}\mathbf F_{\mathrm{RF}}.
	\label{eq:AS_def}
\end{equation}
Assuming that the involved Gram matrix is nonsingular, the digital variables are updated by solving a weighted least-squares problem, which yields
\begin{equation}
	\mathbf X^{\star}
	=
	(\mathbf A_{\mathcal S}^{\mathrm H}\mathbf A_{\mathcal S})^{-1}
	\mathbf A_{\mathcal S}^{\mathrm H}
	\mathbf L_{\mathcal S}
	[\widetilde{\mathbf W}_{\mathcal S}^{\mathrm{bal}},\,\widetilde{\mathbf v}_{\mathcal S}^{\mathrm{bal}}],
	\label{eq:X_update}
\end{equation}
where \(\mathbf X^{\star}=[\mathbf F_{\mathrm s}^{\star},\,\mathbf f_{\mathrm a}^{\star}]\).

For fixed \((\mathbf F_{\mathrm s},\mathbf f_{\mathrm a})\), define
\begin{equation}
	\bar{\mathbf T}_{\mathcal S}
	\triangleq
	\mathbf L_{\mathcal S}
	[\widetilde{\mathbf W}_{\mathcal S}^{\mathrm{bal}},\,\widetilde{\mathbf v}_{\mathcal S}^{\mathrm{bal}}],
	\qquad
	\mathbf Y
	\triangleq
	\mathbf L_{\mathcal S}\mathbf F_{\mathrm{RF}}\mathbf X.
	\label{eq:Tbar_Y_def}
\end{equation}
Let \(\mathbf f_n\in\mathbb C^{|\mathcal S|\times 1}\) denote the \(n\)-th column of \(\mathbf F_{\mathrm{RF}}\), and let \(\mathbf x_n\triangleq(\mathbf X_{n,:})^{\mathrm H}\in\mathbb C^{(K+1)\times 1}\).
The residual associated with the \(n\)-th RF column is
\begin{equation}
	\mathbf E_n
	\triangleq
	\bar{\mathbf T}_{\mathcal S}
	-
	\mathbf Y
	+
	\mathbf L_{\mathcal S}\mathbf f_n\mathbf x_n^{\mathrm H}.
	\label{eq:En_def}
\end{equation}
For \(\mathbf x_n\neq\mathbf 0\), the corresponding unconstrained update direction is
\begin{equation}
	\mathbf g_n
	=
	(\mathbf L_{\mathcal S}^{\mathrm H}\mathbf L_{\mathcal S})^{-1}
	\mathbf L_{\mathcal S}^{\mathrm H}
	\mathbf E_n\mathbf x_n^{\ast}
	\big/
	\|\mathbf x_n\|_2^2.
	\label{eq:gn_def}
\end{equation}
The RF column is then obtained by projecting \(\mathbf g_n\) onto the constant-modulus set:
\begin{equation}
	\mathbf f_n
	\leftarrow
	\frac{1}{\sqrt{|\mathcal S|}}
	\exp\!\bigl(\iota\,\angle(\mathbf g_n)\bigr).
	\label{eq:fn_projection}
\end{equation}

After each alternating update, the BF and AN branches are separately rescaled to match the target powers:
\begin{equation}
	\mathbf F_{\mathrm s}
	\leftarrow
	\eta_{\mathrm s}\mathbf F_{\mathrm s},
	\qquad
	\mathbf f_{\mathrm a}
	\leftarrow
	\eta_{\mathrm a}\mathbf f_{\mathrm a},
	\label{eq:rescaling_step}
\end{equation}
where
\begin{equation}
	\eta_{\mathrm s}
	=
	\sqrt{
	\frac{P_{\mathrm{BF}}^{\star}}
	{\operatorname{Tr}(\mathbf W_{\mathrm{HB}}\mathbf W_{\mathrm{HB}}^{\mathrm H})}
	},
	\qquad
	\eta_{\mathrm a}
	=
	\sqrt{
		\frac{P_{\mathrm{AN}}^{\star}}
		{\|\mathbf v_{\mathrm{HB}}\|_2^2}
	}.
	\label{eq:etas_def}
\end{equation}
By alternating among \eqref{eq:X_update}, \eqref{eq:fn_projection}, and \eqref{eq:rescaling_step}, we obtain the hybrid BF--AN pair \((\mathbf W_{\mathrm{HB}},\mathbf v_{\mathrm{HB}})\) on the selected active-port set, which is used for the final transmission and secrecy-rate evaluation.
Alternatively, the RF precoder under the constant-modulus constraints can also be updated by manifold-optimization-based methods~\cite{Zhang2019BeamSequence,Ma2023RISCellFree}.

\vspace{-0.5cm}

\subsection{Algorithm Overview and Complexity Analysis}
\label{subsec:algo_complexity}

The proposed method follows an AO structure with a continuous BF/AN refit step and a discrete active-port pruning step. Starting from an initial support set $\mathcal S_0$ with $|\mathcal S_0|=L$, the algorithm first updates the fully digital BF/AN variables on the current support according to \eqref{eq:secrecy_rate_reformulated}--\eqref{eq:balanced_updates}. It then computes the row-energy score according to \eqref{eq:row_energy_def}, determines the pruning batch size according to \eqref{eq:batch_size}, and updates the deletion set, support, and selection matrix according to \eqref{eq:delete_set}--\eqref{eq:P_update_min}. These staged prune--refit iterations continue until the support cardinality is reduced to $N_{\mathrm t}$. Finally, a long refit is performed on the terminal support, followed by the HBF realization according to \eqref{eq:hbf_bf_an}--\eqref{eq:etas_def}. The complete procedure is summarized in Algorithm~\ref{alg:ao_prune_refit}.

\begin{algorithm}[!t]

\caption{AO framework with staged prune--refit for AN-aided NF FA--MIMO secure transmission}
\label{alg:ao_prune_refit}
\begin{algorithmic}[1]

\Require noise-whitened equivalent channels $\bar{\mathbf H}_{\mathrm U}$ and $\bar{\mathbf H}_{\mathrm E}$; total transmit power budget $P_{\mathrm t}$; number of data streams $K$; number of candidate FA ports $L$; number of active transmit ports $N_{\mathrm t}$; number of RF chains $N_{\mathrm{RF}}$; initial support set $\mathcal S_0$; pruning ratio $\eta$; minimum pruning batch size $m_{\min}$
\Ensure final active-port support $\mathcal S^\star$; balanced fully digital BF matrix $\widetilde{\mathbf W}_{\mathcal S^\star}^{\mathrm{bal}}$; balanced fully digital AN vector $\widetilde{\mathbf v}_{\mathcal S^\star}^{\mathrm{bal}}$; hybrid BF matrix $\mathbf W_{\mathrm{HB}}$; hybrid AN vector $\mathbf v_{\mathrm{HB}}$
\State Initialize $\mathcal S \leftarrow \mathcal S_0$
\While{$|\mathcal S| > N_{\mathrm t}$}
    \State Update the fully digital BF/AN variables according to \eqref{eq:secrecy_rate_reformulated}--\eqref{eq:balanced_updates}
    \State Compute the row-energy score according to \eqref{eq:row_energy_def}
    \State Determine the pruning batch size according to \eqref{eq:batch_size}
    \State Update the deletion set, support, and selection matrix according to \eqref{eq:delete_set}--\eqref{eq:P_update_min}
\EndWhile
\State Perform a final fully digital refit on $\mathcal S^\star$ according to \eqref{eq:secrecy_rate_reformulated}--\eqref{eq:balanced_updates}
\State Construct the HBF realization according to \eqref{eq:hbf_bf_an}--\eqref{eq:etas_def}
\end{algorithmic}
\end{algorithm}

The complexity is dominated by the repeated fully digital refits in the staged prune--refit procedure.
Let \(M_q \triangleq |\mathcal S_q|\) denote the support size at pruning stage \(q\), \(I_{\mathrm{BCD}}\) the number of BCD iterations, \(I_{\lambda}\) the number of bisection steps used in solving \eqref{eq:waterfill_eq_gswf}, \(T_{\mathrm{pr}}\) the number of pruning stages, and \(I_{\mathrm{HBF}}\) the number of iterations in the final HBF fitting stage.
For a given support of size \(M_q\), one fully digital refit is dominated by matrix multiplications, matrix inversions, and eigendecompositions on \(M_q\times M_q\) matrices, which yields
\begin{equation}
\mathcal C_{\mathrm{FD}}(M_q)
=
\mathcal{O}\!\left(
I_{\mathrm{BCD}} M_q^3
+
I_{\lambda} M_q (K+1)
\right).
\end{equation}
The pruning step only incurs a lower-order cost due to row-norm evaluation and ranking, namely
\begin{equation}
\mathcal C_{\mathrm{prune}}(M_q)
=
\mathcal{O}\!\left(
M_q (K+1)
+
M_q \log M_q
\right).
\end{equation}
After the terminal support is reached, the final HBF realization on \(|\mathcal S^\star|=N_{\mathrm t}\) requires alternating least-squares and phase-projection updates, whose complexity is
\begin{equation}
\mathcal C_{\mathrm{HBF}}
=
\mathcal{O}\!\left(
I_{\mathrm{HBF}}\, N_{\mathrm t} N_{\mathrm{RF}} (K+1)
+
I_{\mathrm{HBF}}\, N_{\mathrm{RF}}^3
\right).
\end{equation}
Hence, the total complexity can be written as
\begin{equation}
\mathcal C_{\mathrm{total}}
=
\sum_{q=0}^{T_{\mathrm{pr}}-1}
\left(
\mathcal C_{\mathrm{FD}}(M_q)
+
\mathcal C_{\mathrm{prune}}(M_q)
\right)
+
\mathcal C_{\mathrm{HBF}}.
\end{equation}
Using the bound \(M_q \le L\) for all stages, a coarse asymptotic upper bound is
\begin{equation}
\begin{aligned}
\mathcal C_{\mathrm{total}}
=
\mathcal{O}\!\Big(
&\,T_{\mathrm{pr}} I_{\mathrm{BCD}} L^3
+ T_{\mathrm{pr}} L \log L \\
&+ I_{\mathrm{HBF}} N_{\mathrm{RF}}
\big(N_{\mathrm t}(K+1)+N_{\mathrm{RF}}^2\big)
\Big).
\end{aligned}
\end{equation}
Therefore, the proposed method has polynomial-time complexity, and its dominant growth is governed by the repeated cubic-cost fully digital refits over the intermediate supports.

\section{Numerical Results and Discussion}
\label{sec:numerical_results}
Numerical results are presented to characterize the operating regimes of the proposed AN-aided secure NF FA--MIMO design under different aperture sizes, transmit-power regions, propagation geometries, channel conditions, and implementation settings. Uniform linear arrays (ULAs) are deployed at the BS (Alice), the legitimate user (Bob), and the potential eavesdropper (Eve), with their array-center boresights aligned on a common line. Unless otherwise stated, the geometric configuration follows \cite{zhang2024physical} and is illustrated in Fig.~\ref{coordination_figure}: Bob and Eve are located at the same azimuth angle of \(45^{\circ}\), equipped with FPA arrays of sizes \(N_{\mathrm u}=8\) and \(N_{\mathrm e}=8\), and placed at radial distances \(d_{\mathrm B}=15\,\mathrm{m}\) and \(d_{\mathrm E}=5\,\mathrm{m}\) from the base-station array center. At Alice, the FPA baseline adopts half-wavelength spacing, so its transmit aperture is \(D_{\mathrm A}=(N_{\mathrm t}-1)\lambda/2\). For a fair comparison, the FA rail is configured to have the same overall aperture, i.e., \((L-1)d_{\mathrm{port}}=D_{\mathrm A}\). Bob and Eve use half-wavelength FPA arrays, with \(D_{\mathrm B}=(N_{\mathrm u}-1)\lambda/2\) and \(D_{\mathrm E}=(N_{\mathrm e}-1)\lambda/2\). To relate the numerical setup to practical circuit-controlled FAS hardware, we consider a pixel-based FAS implementation, where the antenna-state reconfiguration time is at the microsecond level under FPGA-based electronic control~\cite{zhang2024PRA}. In addition, the PIN-switch bias for each active port is on the order of \(1.33\,\mathrm{V}\) and \(10\,\mathrm{mA}\), corresponding to a switching-control power of about \(13.3\,\mathrm{mW}\) per active switch. Hence, compared with mechanical or liquid-driven implementations, the practical reconfiguration overhead is mainly a modest electronic control cost. The noise power is \(\sigma^{2}=-105\,\mathrm{dBm}\), the number of data streams is \(K=4\), and the number of RF chains at the transmitter is \(N_{\mathrm{RF}}=8\). 
For the staged pruning procedure, the pruning ratio and the minimum pruning batch size are set to \(\eta=0.2\) and \(m_{\min}=1\), respectively, throughout all simulations.
All performance metrics are averaged over 100 independent realizations of the beamforming matrices and AN vectors, whose entries are initialized as independent complex Gaussian samples.

\begin{figure}
	\centering
	\includegraphics[width=0.45\textwidth]{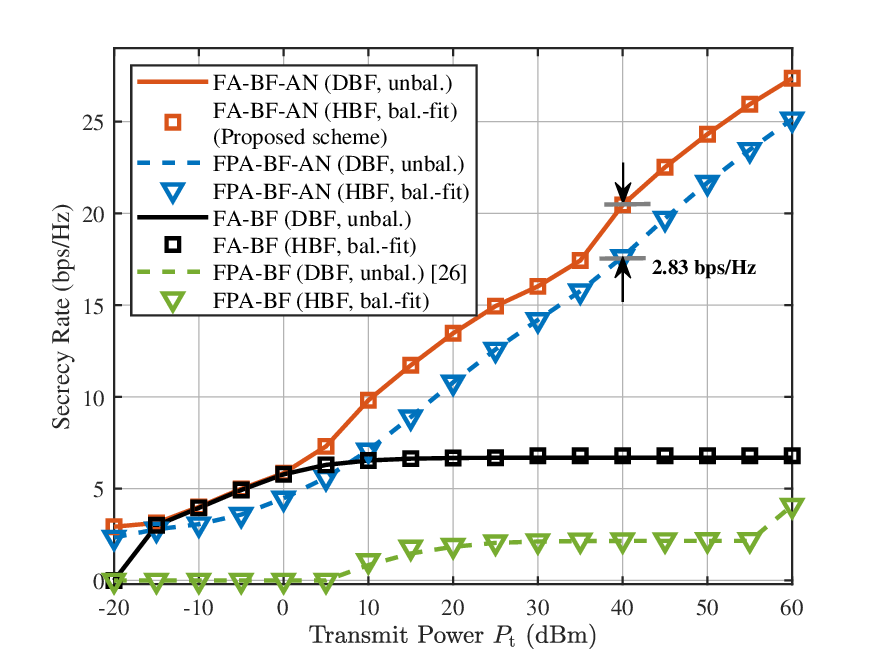}
	
	\caption{ SR for FA–MIMO with \(L=64\) and \(N_{\mathrm t}=16\); comparison with an FPA-based BF-only baseline, with ablations over array type (FA vs. FPA), AN co-design (with/without), and implementation (DBF vs. HBF after power balancing).	\label{fig:small_array_compare}
	}
	
	\vspace{-0.5cm}
\end{figure}

Fig.~\ref{fig:small_array_compare} reports the SR performance of the considered PLS schemes under a small-array setting with \(L=64\) and \(N_{\mathrm t}=16\). Under the above aperture-matching rule, the FA candidate-port spacing is approximately \(d_{\mathrm{port}}=\lambda/8\) in this particular setting. Hence, the larger number of FA candidate ports is realized within the same transmit aperture rather than by enlarging the array size. The carrier frequency is \(f=2.8\,\mathrm{GHz}\), which yields a MIMO Rayleigh distance \(r_{\mathrm{MIMO\text{-}RD}}=2(D_{\mathrm A}+D_{\mathrm B})^{2}/\lambda=25.9\,\mathrm{m}\), so Bob at \(d_{\mathrm B}=15\,\mathrm{m}\) and Eve at \(d_{\mathrm E}=5\,\mathrm{m}\) both lie in the near-field region. The results show that, in this small-array regime, both AN co-design and FA-based port reconfiguration provide clear secrecy gains. For example, at \(P_{\mathrm t}=40\,\mathrm{dBm}\), the proposed \(\mathrm{FA}\text{-}\mathrm{BF}\text{-}\mathrm{AN}\) scheme achieves \(20.46\,\mathrm{bps/Hz}\), whereas \(\mathrm{FPA}\text{-}\mathrm{BF}\text{-}\mathrm{AN}\) attains \(17.63\,\mathrm{bps/Hz}\), yielding an FA gain of \(2.83\,\mathrm{bps/Hz}\). In contrast, the BF-only baselines are much weaker: at the same power, \(\mathrm{FA}\text{-}\mathrm{BF}\) and \(\mathrm{FPA}\text{-}\mathrm{BF}\) achieve only about \(6.79\,\mathrm{bps/Hz}\) and \(2.20\,\mathrm{bps/Hz}\), respectively, which confirms that AN plays the dominant role in this unfavorable near-field geometry with \(d_{\mathrm E}<d_{\mathrm B}\). Moreover, the BF-only curves exhibit an evident saturation effect, while the BF+AN curves continue to increase with \(P_{\mathrm t}\). At \(P_{\mathrm t}=60\,\mathrm{dBm}\), \(\mathrm{FA}\text{-}\mathrm{BF}\text{-}\mathrm{AN}\) still maintains a \(2.23\,\mathrm{bps/Hz}\) advantage over \(\mathrm{FPA}\text{-}\mathrm{BF}\text{-}\mathrm{AN}\), showing that the FA gain is persistent rather than incidental. This trend is also reflected in the optimized power allocation of the proposed FA-BF-AN scheme. At \(P_{\mathrm t}=60\,\mathrm{dBm}\), the optimizer yields \((P_W,P_v)=(7.51\times10^5,\,2.49\times10^5)\), meaning that about \(75.1\%\) of the total transmit power is assigned to BF and \(24.9\%\) to AN. Finally, the HBF curves almost overlap with the DBF references, indicating that the proposed hybrid realization preserves nearly all of the digital-domain secrecy gain. Overall, Fig.~\ref{fig:small_array_compare} suggests that, for compact apertures, secrecy enhancement mainly comes from AN transmission, while FA further provides a stable additional gain through active-port reconfiguration.

\begin{figure}
	\centering
	\includegraphics[width=0.45\textwidth]{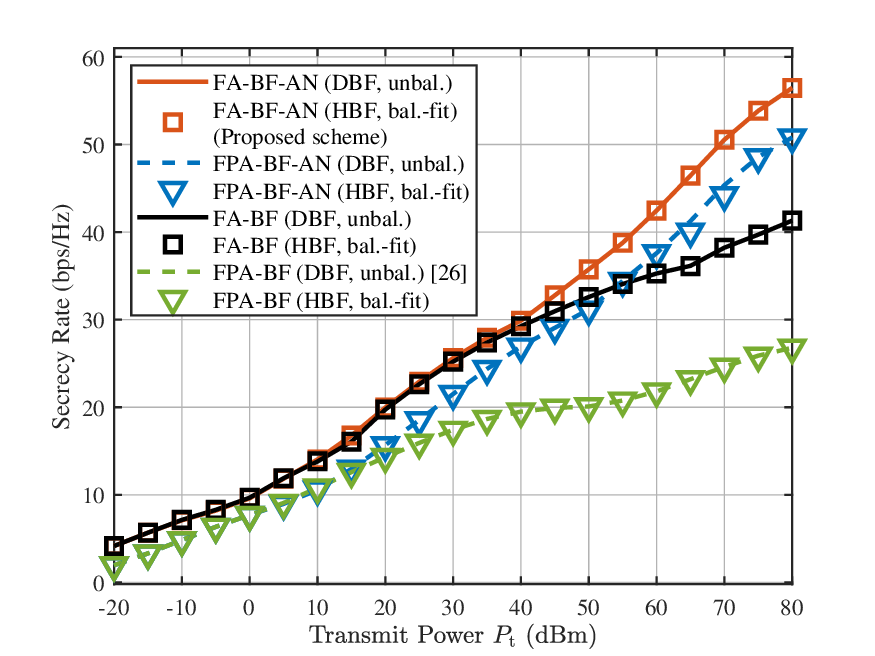}
	
	\caption{ SR for FA–MIMO with \(L=512\) and \(N_{\mathrm t}=128\); comparison with an FPA-based BF-only baseline, with ablations over array type (FA vs. FPA), AN co-design (with/without), and implementation (DBF vs. HBF after power balancing).\label{fig:large_array_compare}
	}
	
	\vspace{-0.5cm}
\end{figure}

Fig.~\ref{fig:large_array_compare} shows the SR performance in the large-array case with \(L=512\) and \(N_{\mathrm t}=128\). Under the same aperture-matching rule, the FA candidate-port spacing also becomes \(d_{\mathrm{port}}=\lambda/8\) in this particular setting. At \(f=28\,\mathrm{GHz}\), this corresponds to \(r_{\mathrm{MIMO\text{-}RD}}=2(D_{\mathrm A}+D_{\mathrm B})^{2}/\lambda\approx96.1\,\mathrm{m}\). Compared with Fig.~\ref{fig:small_array_compare}, the dominant source of gain becomes power-dependent. In the low-\(P_{\mathrm t}\) region, the improvement mainly comes from FA. For example, at \(P_{\mathrm t}=10\,\mathrm{dBm}\), the HBF-based \(\mathrm{FA}\text{-}\mathrm{BF}\) scheme achieves \(13.82\,\mathrm{bps/Hz}\), whereas \(\mathrm{FPA}\text{-}\mathrm{BF}\) attains \(10.84\,\mathrm{bps/Hz}\), giving an FA gain of about \(2.99\,\mathrm{bps/Hz}\). The power-allocation results of the proposed FA-BF-AN scheme show that, in the low- and moderate-power region, the optimized AN power is indeed negligible: for \(P_{\mathrm t}=-15\), \(10\), and \(30\,\mathrm{dBm}\), the reported solution satisfies \(P_v\approx0\), so essentially all transmit power is assigned to BF. As \(P_{\mathrm t}\) increases, the advantage of AN becomes more visible. In particular, at \(P_{\mathrm t}=70\,\mathrm{dBm}\), the proposed \(\mathrm{FA}\text{-}\mathrm{BF}\text{-}\mathrm{AN}\) scheme reaches \(50.55\,\mathrm{bps/Hz}\), which is \(12.34\,\mathrm{bps/Hz}\) higher than \(\mathrm{FA}\text{-}\mathrm{BF}\) and \(25.92\,\mathrm{bps/Hz}\) higher than \(\mathrm{FPA}\text{-}\mathrm{BF}\). More importantly, the optimized power split indicates that AN is no longer negligible in the high-power region. At \(P_{\mathrm t}=65\,\mathrm{dBm}\), the proposed FA-BF-AN solution uses \((P_W,P_v)=(2.51\times10^6,\,6.57\times10^5)\), corresponding to about \(79.2\%\) of the total power for BF and \(20.8\%\) for AN. At \(P_{\mathrm t}=75\,\mathrm{dBm}\), this becomes \((P_W,P_v)=(2.09\times10^7,\,1.07\times10^7)\), i.e., about \(66.2\%\) for BF and \(33.8\%\) for AN. Therefore, in the large-array regime, AN is negligible only in the low- and moderate-power region, whereas in the high-power region a non-negligible fraction of the transmit power is again allocated to AN. The reason is that the low-power region is mainly noise-limited, so transmit power is more effectively used to strengthen Bob's desired signal, whereas in the high-power region the beamforming gain has already been largely exploited and allocating part of the extra power to AN becomes more effective in further suppressing Eve. Moreover, the HBF curves remain close to the DBF references, confirming that the proposed hybrid implementation preserves the main digital-domain secrecy gains.

\begin{figure}[!t]
	\centering
	\subfloat[]{%
		\includegraphics[width=0.4\textwidth]{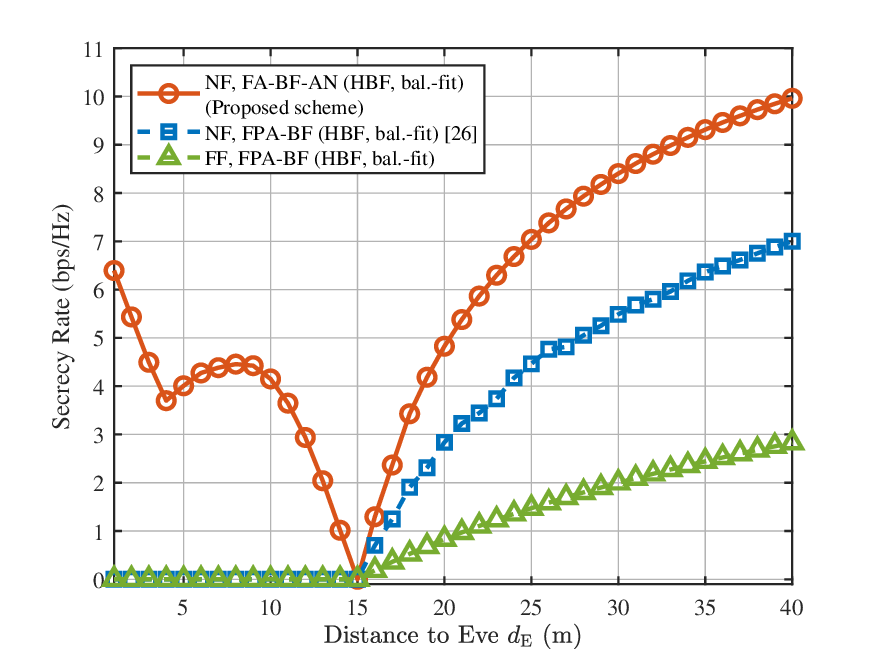}%
		\label{subfig:range_small}}\\[-1pt]
	\subfloat[]{%
		\includegraphics[width=0.4\textwidth]{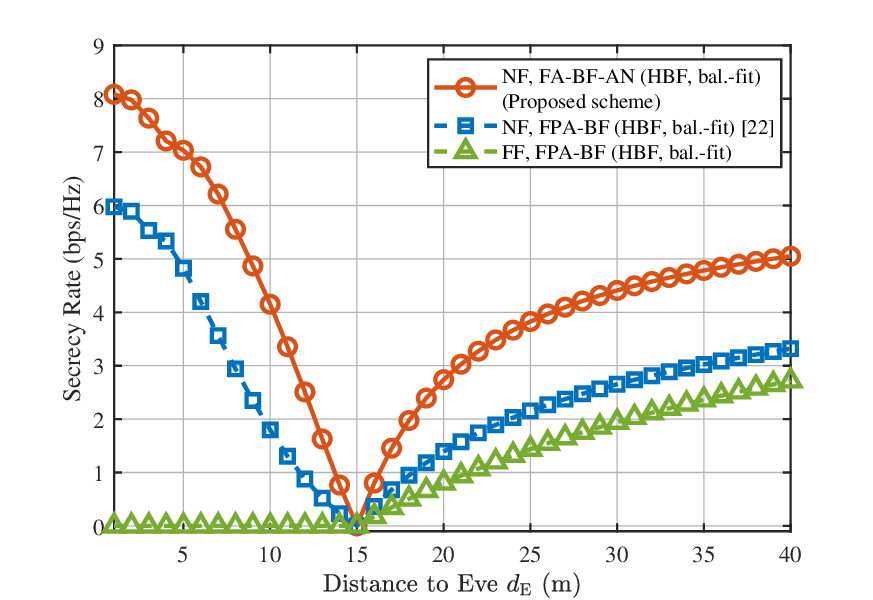}%
		\label{subfig:range_large}}
	\caption{Secrecy performance of PLS schemes with Bob fixed at \(d_{\mathrm B}=15\,\mathrm{m}\) and transmit power \(P_{\mathrm t}=-10\,\mathrm{dBm}\), evaluated over different distances between Alice and Eve: (a) small array (\(L=64\), \(N_{\mathrm t}=16\)); (b) large array (\(L=512\), \(N_{\mathrm t}=128\)).\label{fig:range_compare}}
	\vspace{-0.5cm}
\end{figure}

Fig.~\ref{fig:range_compare} shows the secrecy-rate variation versus Eve distance \(d_{\mathrm E}\) with Bob fixed at \(d_{\mathrm B}=15\,\mathrm{m}\) and transmit power \(P_{\mathrm t}=-10\,\mathrm{dBm}\). The compared schemes are the proposed near-field (NF) FA-BF-AN scheme, the NF FPA-BF baseline, and the far-field (FF) FPA-BF baseline. For a fair comparison, the FF scheme uses the same large-scale path-loss scaling as the NF scheme, while its small-scale fading adopts a far-field model.

In Fig.~\ref{fig:range_compare}\subref{subfig:range_small}, corresponding to the small-array setting \((L,N_{\mathrm t})=(64,16)\) and \(f=2.8\,\mathrm{GHz}\), the proposed NF FA-BF-AN scheme exhibits a clear non-monotonic trend: the secrecy rate decreases in the very near region, partially recovers, drops again to an almost zero minimum around \(d_{\mathrm E}\approx d_{\mathrm B}\), and then increases for \(d_{\mathrm E}>d_{\mathrm B}\). This indicates that, with a limited aperture, Bob and Eve cannot be well separated when their radial positions are close, so the effective channel overlap remains strong and the AN leakage toward Bob becomes non-negligible. By contrast, the two BF-only baselines stay close to zero for \(d_{\mathrm E}\le d_{\mathrm B}\) and improve only when Eve moves farther away, while the NF FPA-BF baseline remains above the FF FPA-BF baseline in the large-\(d_{\mathrm E}\) region.

In Fig.~\ref{fig:range_compare}\subref{subfig:range_large}, corresponding to the large-array setting \((L,N_{\mathrm t})=(512,128)\) and \(f=28\,\mathrm{GHz}\), the same overall trend is preserved, while the separation among the three schemes becomes more evident over a wider distance range. Under this setting, the proposed NF FA-BF-AN scheme achieves the highest secrecy rate over almost the entire distance range except near \(d_{\mathrm E}\approx d_{\mathrm B}\), where all schemes approach zero due to the strongest Bob--Eve channel coupling. The NF FPA-BF baseline is again consistently superior to the FF FPA-BF baseline, indicating that the near-field model retains a clearer spatial selectivity advantage in this regime, while the proposed FA-assisted BF-AN co-design further translates this advantage into additional secrecy-rate improvement.

\begin{figure}[!t]
	\centering
	\subfloat[]{%
		\includegraphics[width=0.5\linewidth]{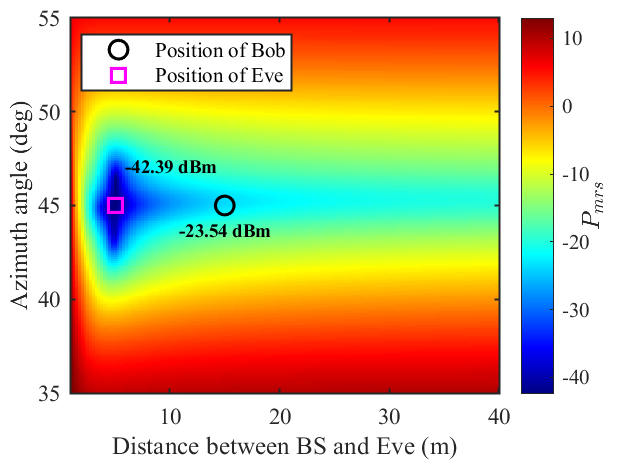}%
		\label{subfig:rx_bf_small}}
	\hfill
	\subfloat[]{%
		\includegraphics[width=0.5\linewidth]{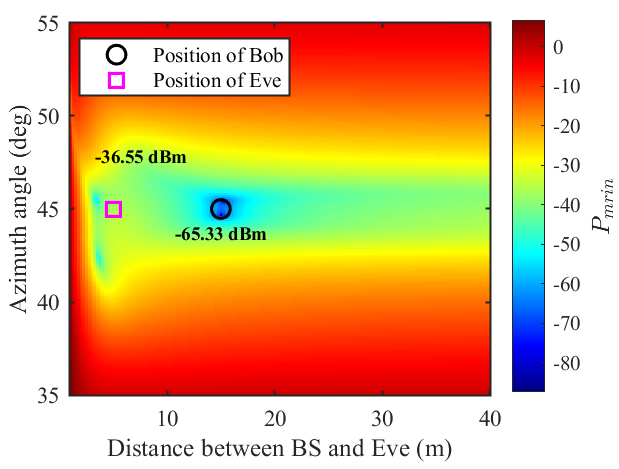}%
		\label{subfig:rx_an_small}}
	\\[-6pt]
	\subfloat[]{%
		\includegraphics[width=0.5\linewidth]{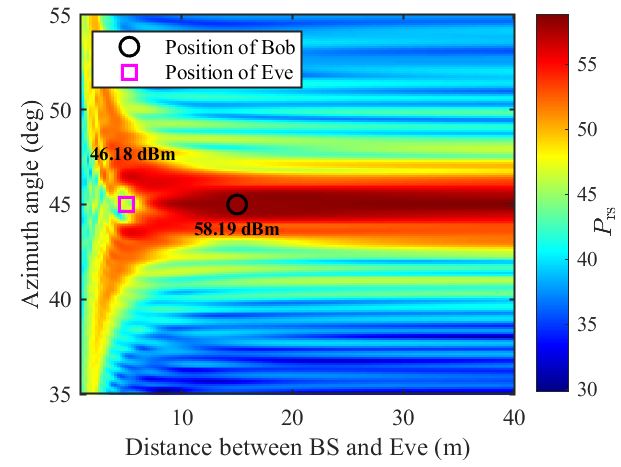}%
		\label{subfig:rx_bf_large}}
	\hfill
	\subfloat[]{%
		\includegraphics[width=0.5\linewidth]{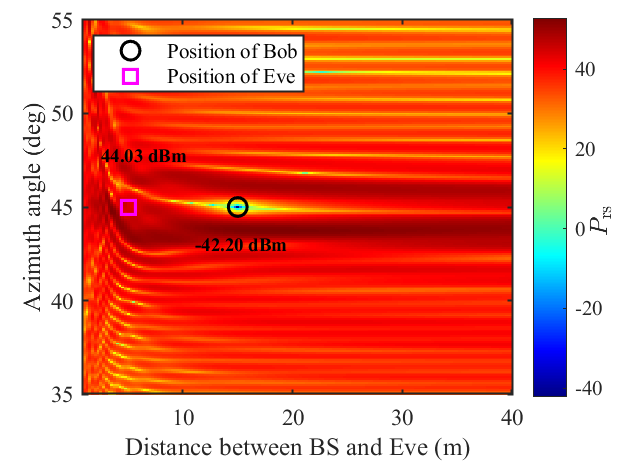}%
		\label{subfig:rx_an_large}}

	\caption{Received signal power and interference-plus-noise power distributions of the proposed FA-BF-AN scheme: (a) small-array received signal power heatmap at \(f=2.8\,\mathrm{GHz}\), \((L,N_{\mathrm t})=(64,16)\), and \(P_{\mathrm t}=20\,\mathrm{dBm}\); (b) small-array interference-plus-noise power heatmap under the same setting; (c) large-array received signal power heatmap at \(f=28\,\mathrm{GHz}\), \((L,N_{\mathrm t})=(512,256)\), and \(P_{\mathrm t}=65\,\mathrm{dBm}\); (d) large-array interference-plus-noise power heatmap under the same setting.\label{fig:rx_2x2}}
	
	\vspace{-0.5cm}
\end{figure}

Fig.~\ref{fig:rx_2x2} presents the received signal power (RSP) and interference-plus-noise power (INP) distributions over free-space locations for the proposed FA-BF-AN design. Specifically, Figs.~\ref{fig:rx_2x2}\subref{subfig:rx_bf_small} and \subref{subfig:rx_an_small} correspond to the small-array setting with \(f=2.8\,\mathrm{GHz}\), \((L,N_{\mathrm t})=(64,16)\), and \(P_{\mathrm t}=20\,\mathrm{dBm}\), whereas Figs.~\ref{fig:rx_2x2}\subref{subfig:rx_bf_large} and \subref{subfig:rx_an_large} correspond to the large-array setting with \(f=28\,\mathrm{GHz}\), \((L,N_{\mathrm t})=(512,256)\), and \(P_{\mathrm t}=65\,\mathrm{dBm}\). Hence, the absolute power levels in the small- and large-array cases are not intended for direct comparison; instead, the figure is used to highlight the spatial power distributions and the relative power separation between Bob and Eve. In the small-array case, Fig.~\ref{fig:rx_2x2}\subref{subfig:rx_bf_small} shows that the received signal power at Bob is \(-23.54\,\mathrm{dBm}\), while that at Eve is reduced to \(-42.39\,\mathrm{dBm}\), indicating that BF already provides a noticeable signal-power advantage to Bob. In Fig.~\ref{fig:rx_2x2}\subref{subfig:rx_an_small}, the INP at Bob is \(-65.33\,\mathrm{dBm}\), whereas the corresponding value at Eve is \(-36.55\,\mathrm{dBm}\), which shows that the AN component is mainly concentrated around Eve while remaining much weaker at Bob. Therefore, under a compact aperture, secrecy enhancement relies on the joint effect of BF and AN, where BF provides the desired-signal advantage and AN further enlarges the SINR gap between Bob and Eve. In the large-array case, Fig.~\ref{fig:rx_2x2}\subref{subfig:rx_bf_large} shows a much sharper and more concentrated mainlobe toward Bob, with received signal powers of \(58.19\,\mathrm{dBm}\) at Bob and \(46.18\,\mathrm{dBm}\) at Eve, indicating stronger spatial focusing and better location selectivity. In Fig.~\ref{fig:rx_2x2}\subref{subfig:rx_an_large}, the INP at Bob is \(-42.20\,\mathrm{dBm}\), while the corresponding value at Eve reaches \(44.03\,\mathrm{dBm}\), which confirms that, with the enlarged aperture and higher carrier frequency, the AN can be much more effectively directed toward Eve-dominant spatial regions while being strongly suppressed at Bob. These results verify that, as the array aperture increases, the near-field spatial selectivity becomes significantly stronger, and the proposed FA-assisted BF-AN co-design can exploit this higher spatial resolution to achieve a much clearer power separation between the legitimate and eavesdropping locations. 

\begin{figure}[!t]
	\centering
	\subfloat[]{%
		\includegraphics[width=0.8\linewidth]{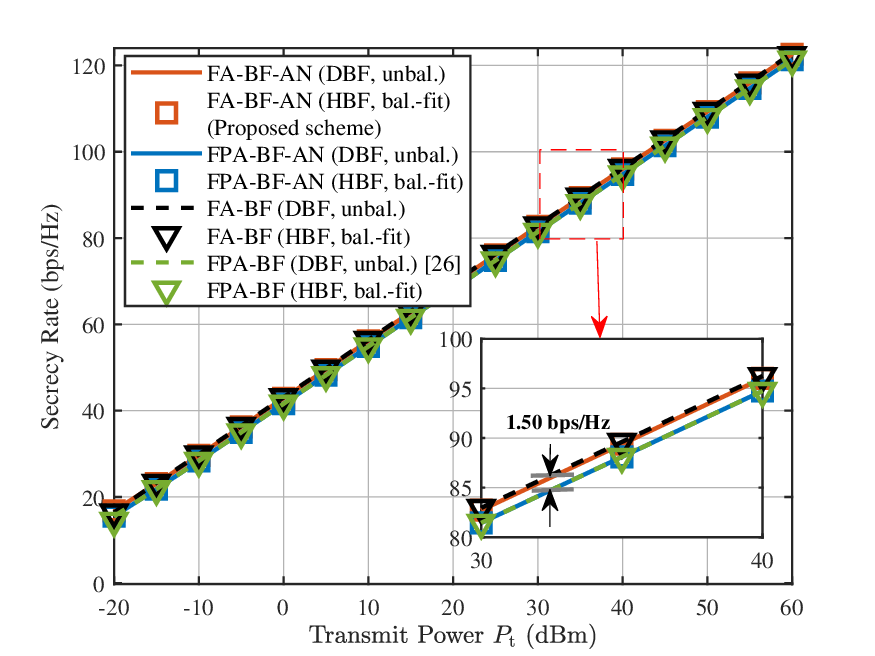}%
		\label{subfig:rician_perfect}}\\[-1pt]
	\subfloat[]{%
		\includegraphics[width=0.8\linewidth]{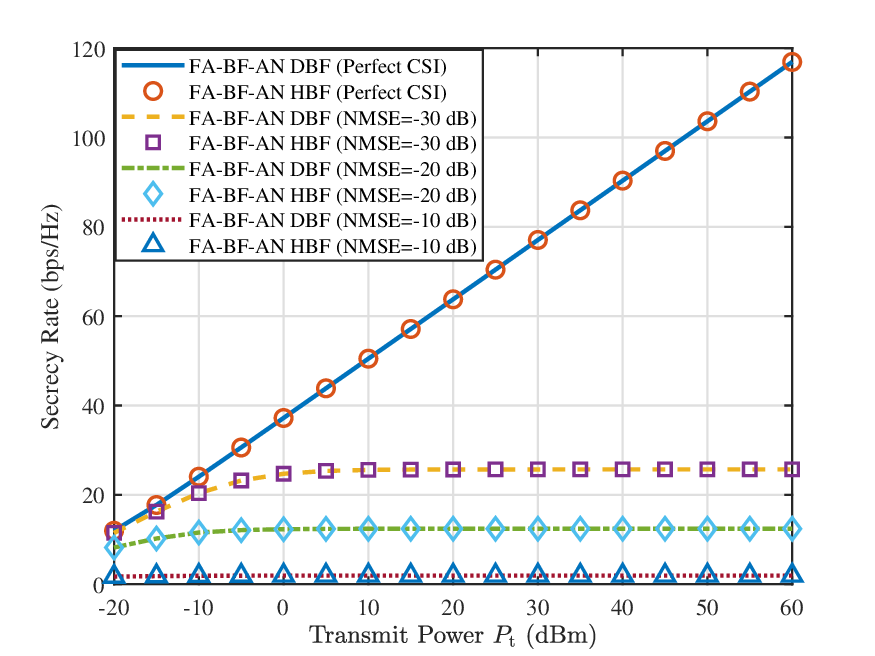}%
		\label{subfig:rician_imperfect}}
	
	\caption{Secrecy performance of the proposed FA-BF-AN scheme under Rician fading: (a) perfect CSI with \(f=2.8\,\mathrm{GHz}\), \(L=64\), \(N_{\mathrm t}=16\), \(\sigma^2=-105\,\mathrm{dBm}\), and \(K_{\mathrm B}=K_{\mathrm E}=-5\,\mathrm{dB}\); (b) imperfect Eve CSI with \(f=2.8\,\mathrm{GHz}\), \(L=64\), \(N_{\mathrm t}=32\), \(\sigma^2=-105\,\mathrm{dBm}\), and \(K_{\mathrm B}=K_{\mathrm E}=10\,\mathrm{dB}\). In (b), the transmitter designs the scheme based on the estimated Eve channel and evaluates the secrecy rate on the true channel.}
	\vspace{-0.5cm}
	\label{fig:rician_imperfect}
	
\end{figure}

Fig.~\ref{fig:rician_imperfect} shows the secrecy performance of the proposed FA-BF-AN scheme under Rician fading with perfect and imperfect CSI. In Fig.~\ref{fig:rician_imperfect}\subref{subfig:rician_perfect}, the simulation is conducted with \(f=2.8\,\mathrm{GHz}\), \(L=64\) candidate ports, \(N_{\mathrm t}=16\) active ports, and noise power \(\sigma^2=-105\,\mathrm{dBm}\), under the Rician setting \(K_{\mathrm B}=K_{\mathrm E}=-5\,\mathrm{dB}\). The results verify that the proposed scheme is not restricted to the pure LoS case, but remains effective when non-LoS components are present. Moreover, when the scattered component becomes stronger, the legitimate and eavesdropping channels become less aligned, which enlarges the optimization space for joint port selection and BF/AN design and can therefore improve the achievable secrecy performance. In Fig.~\ref{fig:rician_imperfect}\subref{subfig:rician_imperfect}, imperfect Eve CSI is further considered with \(f=2.8\,\mathrm{GHz}\), \(L=64\), \(N_{\mathrm t}=32\), and \(\sigma^2=-105\,\mathrm{dBm}\), under the Rician setting \(K_{\mathrm B}=K_{\mathrm E}=10\,\mathrm{dB}\). In this case, the transmitter designs the scheme based on the estimated Eve channel, whereas the final secrecy rate is evaluated on the true channel. As expected, the secrecy rate decreases as the NMSE increases, since the CSI mismatch degrades the accuracy of port selection and BF/AN/HBF design. Nevertheless, the proposed scheme still maintains a nonzero secrecy advantage over a wide transmit-power range, which demonstrates a reasonable robustness to CSI uncertainty.

\begin{figure}[!t]
	\centering
	\includegraphics[width=0.9\linewidth]{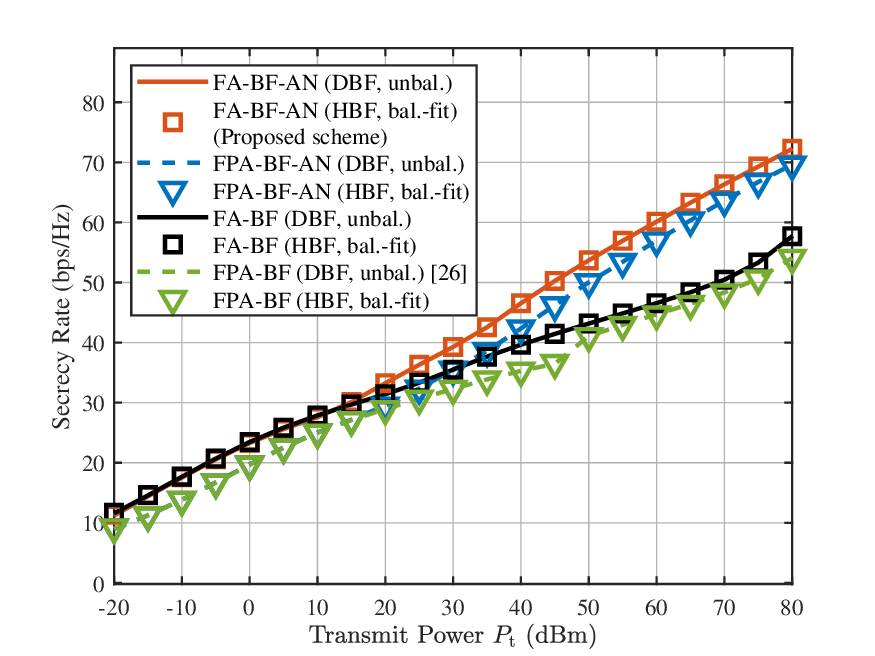}
	
	\caption{Secrecy performance in a different-angle scenario with \(f=2.8\,\mathrm{GHz}\), \(L=64\), \(N_{\mathrm t}=16\), \(N_{\mathrm u}=N_{\mathrm e}=8\), \(K=4\), \(r_{\mathrm A}=1\), \(N_{\mathrm{RF}}=8\), \(\sigma^2=-105\,\mathrm{dBm}\), \(d_{\mathrm B}=d_{\mathrm E}=15\,\mathrm{m}\), \(\theta_{\mathrm B}=45^\circ\), and \(\theta_{\mathrm E}=35^\circ\).}
	\vspace{-0.5cm}
	\label{fig:different_angle}
	
\end{figure}

Fig.~\ref{fig:different_angle} further examines a different-angle scenario, where Bob and Eve are located at the same distance but different azimuth angles. Specifically, the simulation is conducted with \(f=2.8\,\mathrm{GHz}\), \(L=64\) candidate ports, \(N_{\mathrm t}=16\) active ports, \(N_{\mathrm u}=N_{\mathrm e}=8\), \(K=4\), \(r_{\mathrm A}=1\), \(N_{\mathrm{RF}}=8\), and noise power \(\sigma^2=-105\,\mathrm{dBm}\). Bob and Eve are placed at \(d_{\mathrm B}=d_{\mathrm E}=15\,\mathrm{m}\), while their azimuth angles are set to \(\theta_{\mathrm B}=45^\circ\) and \(\theta_{\mathrm E}=35^\circ\), respectively. This result clarifies that the same-angle setting adopted in the main part of the paper is not a simplifying assumption, but rather a deliberately challenging case, since the Bob--Eve distinguishability then mainly relies on the relatively weak radial-domain separation. By contrast, when angular separation is available, the Bob--Eve channel correlation is further reduced, and the secrecy advantage of the proposed design becomes more evident. As shown in Fig.~\ref{fig:different_angle}, the proposed FA-BF-AN scheme achieves the best secrecy performance in the low-to-high transmit-power region, while the HBF realization closely follows its DBF counterpart. For example, at \(P_{\mathrm t}=80\,\mathrm{dBm}\), the proposed FA-BF-AN achieves \(72.24\,\mathrm{bps/Hz}\) with DBF and \(72.24\,\mathrm{bps/Hz}\) with HBF. In comparison, FPA-BF-AN achieves \(69.61\,\mathrm{bps/Hz}\) for both DBF and HBF, FA-BF achieves \(57.71\,\mathrm{bps/Hz}\) for both DBF and HBF, and FPA-BF achieves \(54.07\,\mathrm{bps/Hz}\) with DBF and \(54.08\,\mathrm{bps/Hz}\) with HBF. Hence, at this operating point, the proposed FA-BF-AN provides gains of \(2.63\,\mathrm{bps/Hz}\), \(14.53\,\mathrm{bps/Hz}\), and \(18.17\,\mathrm{bps/Hz}\) over FPA-BF-AN, FA-BF, and FPA-BF, respectively. These results indicate that the proposed scheme is not limited to secrecy enhancement through distance-domain discrimination; instead, it can also exploit angular-domain separability, where FA-enabled port reconfiguration and joint BF-AN design provide additional spatial degrees of freedom for suppressing information leakage.

\begin{figure}[!t]
	\centering
	\includegraphics[width=0.92\linewidth]{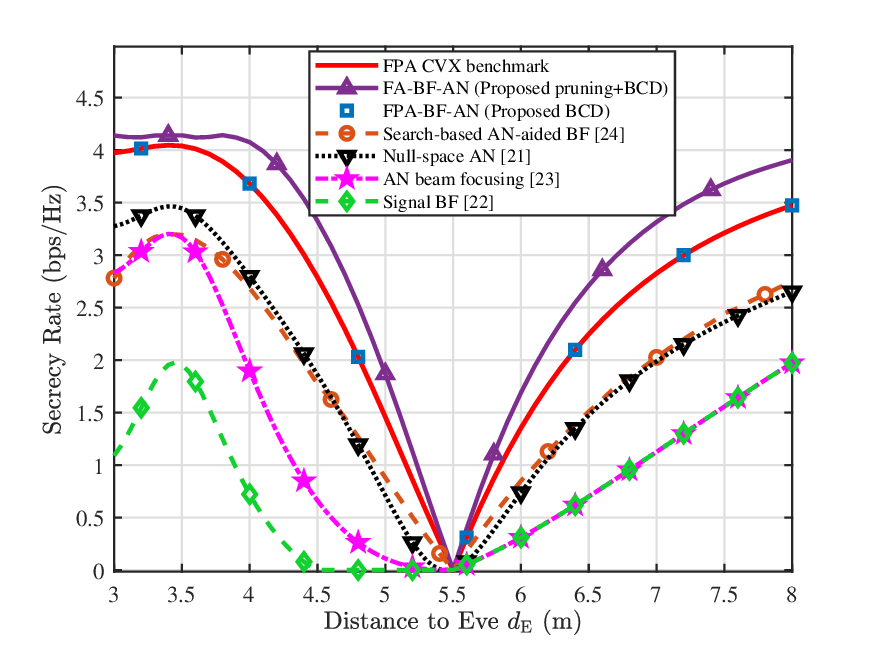}
	
	\caption{Benchmark comparison: secrecy rate versus the eavesdropper distance \(d_{\mathrm E}\), with \(N_{\mathrm{FPA}}=513\), \(f=300~\mathrm{GHz}\), \(d=\lambda/2\), \(P_{\mathrm t}=5~\mathrm{dBm}\), \(\sigma^2=-77~\mathrm{dBm}\), \(\theta_{\mathrm B}=\theta_{\mathrm E}=0\), \(d_{\mathrm B}=5~\mathrm{m}\), \(K(f)=0.00143~\mathrm{m}^{-1}\), and \(M_{\mathrm B}=M_{\mathrm E}=2\).}
	\vspace{-0.5cm}
	\label{fig:benchmark_tang_fig4}

\end{figure}

Fig.~\ref{fig:benchmark_tang_fig4} compares the proposed schemes with representative near-field PLS designs under the simulation setting in \cite{Tang2025ANBF}. To be consistent with our notation, this benchmark corresponds to an FPA-based MISO setting with \(N_{\mathrm u}=N_{\mathrm e}=1\) and \(N_{\mathrm{FPA}}=513\) fixed transmit antennas at Alice, where \(N_{\mathrm{FPA}}\) denotes the number of transmit antennas in the benchmark FPA. The other parameters are set as \(f=300~\mathrm{GHz}\), \(d=\lambda/2\), \(P_{\mathrm t}=5~\mathrm{dBm}\), \(\sigma^2=-77~\mathrm{dBm}\) for a \(5~\mathrm{GHz}\) bandwidth, \(\theta_{\mathrm B}=\theta_{\mathrm E}=0\), and \(d_{\mathrm B}=5~\mathrm{m}\), while \(d_{\mathrm E}\) varies. Moreover, \(K(f)=0.00143~\mathrm{m}^{-1}\) denotes the molecular absorption coefficient at frequency \(f\). It is observed that, in the MISO case, the proposed FPA-BF-AN curve almost coincides with the SCA/CVX-based optimal FPA benchmark, with a maximum deviation of only \(10^{-5}~\mathrm{bps/Hz}\) over the simulated \(d_{\mathrm E}\) range. This confirms that the proposed design essentially achieves the optimal fixed-array performance. By further exploiting FA reconfiguration, the proposed FA-BF-AN scheme consistently provides additional gains over both FPA-BF-AN and the optimal FPA benchmark. Specifically, the gains are \(0.39\), \(0.41\), \(0.46\), and \(0.43~\mathrm{bps/Hz}\) at \(d_{\mathrm E}=4.0\), \(5.0\), \(6.5\), and \(8.0~\mathrm{m}\), respectively. At \(d_{\mathrm E}=8.0~\mathrm{m}\), the proposed FA-BF-AN achieves \(3.90~\mathrm{bps/Hz}\), while FPA-BF-AN and the optimal FPA benchmark both achieve \(3.47~\mathrm{bps/Hz}\); the search-based AN-aided BF \cite{Tang2025ANBF}, null-space AN \cite{Chen2024NFDM}, AN beam focusing \cite{Zhang2025NFPLS}, and signal BF \cite{Nasir2024MaxMin} schemes achieve \(2.74\), \(2.65\), \(1.97\), and \(1.97~\mathrm{bps/Hz}\), respectively. This advantage is attributed to the joint BF-AN design and the additional position-domain degrees of freedom provided by FA. Moreover, the benchmark schemes are developed for MISO settings, where secrecy design mainly reduces to transmit-side beam shaping and BF/AN power allocation. In contrast, the proposed framework targets NF-MIMO, where the secrecy-rate optimization involves coupled matrix-valued channels, multi-stream log-det terms, active-port selection, and HBF realization. Hence, the near-optimal MISO performance and the additional FA gain demonstrate the effectiveness and scalability of the proposed design.

\begin{figure}[!t]
	\centering
	\includegraphics[width=0.92\linewidth]{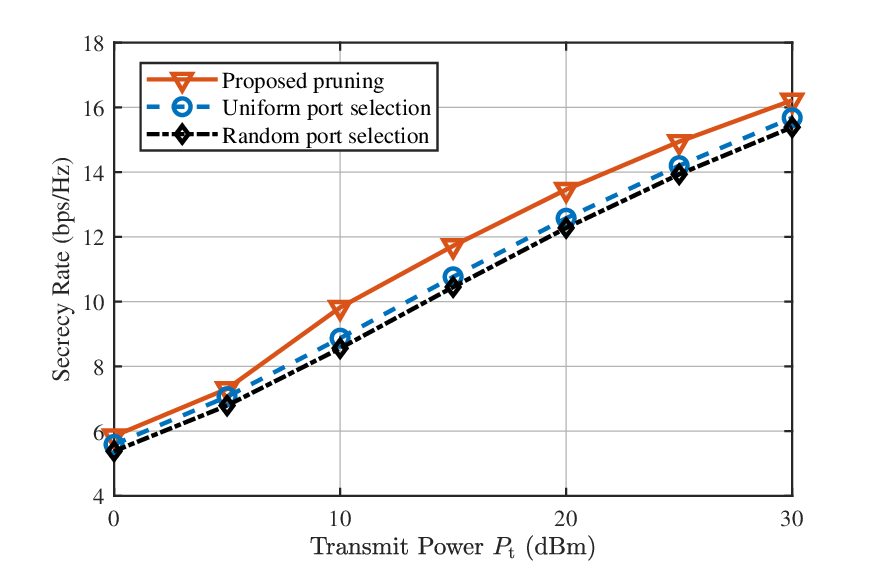}
	
	\caption{Comparison of different port-selection strategies for the FA-BF-AN DBF scheme.}
	\label{fig:port_selection_compare}
	
	\vspace{-0.5cm}
\end{figure}

Fig.~\ref{fig:port_selection_compare} compares the proposed pruning-based port selection with uniform and random port-selection baselines for the FA-BF-AN DBF scheme. The simulation is conducted with \(f=2.8~\mathrm{GHz}\), \(L=64\), \(N_{\mathrm t}=16\), \(N_{\mathrm u}=N_{\mathrm e}=8\), \(K=4\), \(r_{\mathrm A}=1\), \(\sigma^2=-105~\mathrm{dBm}\), \(d_{\mathrm B}=15~\mathrm{m}\), \(d_{\mathrm E}=5~\mathrm{m}\), and \(\theta_{\mathrm B}=\theta_{\mathrm E}=45^\circ\). It is observed that the proposed pruning strategy consistently achieves the highest SR, confirming the effectiveness of channel-aware active-port selection. For example, at \(P_{\mathrm t}=20~\mathrm{dBm}\), the proposed pruning, uniform selection, and random selection achieve \(13.46\), \(12.58\), and \(12.28~\mathrm{bps/Hz}\), respectively, corresponding to gains of \(0.89\) and \(1.19~\mathrm{bps/Hz}\) over the two baselines. This is because the proposed pruning rule adaptively preserves the ports that contribute most to the joint BF-AN design, whereas uniform selection is constrained by a fixed geometric pattern and random selection ignores the channel-dependent port utility.

\section{Conclusion}
\label{conclusion}
This paper investigated secure NF FA--MIMO mobile-access transmission for privacy-sensitive mobile services under a discretized circuit-driven FAS architecture and a finite RF-chain budget. An explicit AO-based design framework was developed to jointly optimize BF/AN signaling, RF-chain-limited active-port reconfiguration, and shared-RF HBF realization in a hardware-consistent manner with explicit computational complexity. The results showed that the proposed design provides meaningful secure-access gains over conventional FPA-based baselines under the same RF-chain budget. In particular, for compact apertures and unfavorable Bob--Eve geometry, AN becomes essential because BF alone is insufficient, whereas for large arrays its benefit is power-dependent and reappears in the high-power region. The results also verified that FA port reconfiguration provides an additional secrecy advantage, the proposed HBF realization closely preserves the fully digital performance, and the overall design remains effective under imperfect Eve CSI and practical circuit-driven FAS constraints. These findings provide regime-dependent design guidance for secure and hardware-constrained mobile communication systems employing emerging fluid-antenna technologies.

\bibliographystyle{IEEEtran}
\bibliography{Reference}

@article{zhang2024PRA,
  author  = {J. Zhang and J. Rao and Z. Li and Z. Ming and C.-Y. Chiu and K.-K. Wong and K.-F. Tong and R. Murch},
  journal = {IEEE Open J. Antennas Propag.},
  title   = {A Novel Pixel-Based Reconfigurable Antenna Applied in Fluid Antenna Systems With High Switching Speed},
  year    = {2025},
  month   = {Feb.},
  volume  = {6},
  number  = {1},
  pages   = {212--228},
  doi     = {10.1109/OJAP.2024.3489215}
}

@article{lu2023Near,
  author  = {Y. Lu and L. Dai},
  journal = {IEEE Trans. Commun.},
  title   = {Near-Field Channel Estimation in Mixed {LoS}/{NLoS} Environments for Extremely Large-Scale {MIMO} Systems},
  year    = {2023},
  month   = {Jun.},
  volume  = {71},
  number  = {6},
  pages   = {3694--3707},
  doi     = {10.1109/TCOMM.2023.3260242}
}

@article{shi2015Secure,
  author  = {Q. Shi and W. Xu and J. Wu and E. Song and Y. Wang},
  journal = {IEEE Trans. Wireless Commun.},
  title   = {Secure Beamforming for {MIMO} Broadcasting With Wireless Information and Power Transfer},
  year    = {2015},
  month   = {May},
  volume  = {14},
  number  = {5},
  pages   = {2841--2853},
  doi     = {10.1109/TWC.2015.2395414}
}

@book{boyd2004convex,
  title     = {Convex Optimization},
  author    = {S. P. Boyd and L. Vandenberghe},
  publisher = {Cambridge Univ. Press},
  year      = {2004}
}

@article{yuan2006model,
  author  = {M. Yuan and Y. Lin},
  journal = {J. Roy. Stat. Soc. B (Statist. Methodol.)},
  title   = {Model Selection and Estimation in Regression With Grouped Variables},
  year    = {2006},
  volume  = {68},
  number  = {1},
  pages   = {49--67},
  month   = {Feb.},
  doi     = {10.1111/j.1467-9868.2005.00532.x}
}

@article{jenatton2011structured,
  author  = {R. Jenatton and J.-Y. Audibert and F. Bach},
  journal = {J. Mach. Learn. Res.},
  title   = {Structured Variable Selection With Sparsity-Inducing Norms},
  year    = {2011},
  volume  = {12},
  pages   = {2777--2824},
  month   = {Nov.}
}

@article{Candes2008ReweightedL1,
  author  = {E. J. Cand{\`e}s and M. B. Wakin and S. P. Boyd},
  journal = {J. Fourier Anal. Appl.},
  title   = {Enhancing Sparsity by Reweighted $\ell_1$ Minimization},
  year    = {2008},
  volume  = {14},
  number  = {5},
  pages   = {877--905},
  month   = {Dec.},
  doi     = {10.1007/s00041-008-9045-x}
}

@article{Blumensath2009IHT,
  author  = {T. Blumensath and M. E. Davies},
  journal = {Appl. Comput. Harmon. Anal.},
  title   = {Iterative Hard Thresholding for Compressed Sensing},
  year    = {2009},
  volume  = {27},
  number  = {3},
  pages   = {265--274},
  month   = {Nov.},
  doi     = {10.1016/j.acha.2009.04.002}
}

@article{zhang2024physical,
  author  = {Z. Zhang and Y. Liu and Z. Wang and X. Mu and J. Chen},
  journal = {IEEE Trans. Veh. Technol.},
  title   = {Physical Layer Security in Near-Field Communications},
  year    = {2024},
  month   = {Jul.},
  volume  = {73},
  number  = {7},
  pages   = {10761--10766},
  doi     = {10.1109/TVT.2024.3366115}
}

@article{Wang2023On,
  author  = {C.-X. Wang and X. You and X. Gao and X. Zhu and Z. Li and et al.},
  journal = {IEEE Commun. Surveys Tuts.},
  title   = {On the Road to {6G}: Visions, Requirements, Key Technologies, and Testbeds},
  year    = {2023},
  month   = {2nd Quart.},
  volume  = {25},
  number  = {2},
  pages   = {905--974},
  doi     = {10.1109/COMST.2023.3249835}
}

@article{Abuhamad2021Sensor,
  author  = {M. Abuhamad and A. Abusnaina and D. Nyang and D. Mohaisen},
  journal = {IEEE Internet Things J.},
  title   = {Sensor-Based Continuous Authentication of Smartphones’ Users Using Behavioral Biometrics: A Contemporary Survey},
  year    = {2021},
  month   = {Jan.},
  volume  = {8},
  number  = {1},
  pages   = {65--84},
  doi     = {10.1109/JIOT.2020.3020076}
}

@article{Leung-Yan-Cheong1978The,
  author  = {S. Leung-Yan-Cheong and M. Hellman},
  journal = {IEEE Trans. Inf. Theory},
  title   = {The {Gaussian} Wire-Tap Channel},
  month   = {Jul.},
  year    = {1978},
  volume  = {24},
  number  = {4},
  pages   = {451--456},
  doi     = {10.1109/TIT.1978.1055917}
}

@article{Yue2025Power,
  author  = {H. Yue and C. Guo and Q. Li and H. Chen and Q. Zhang},
  journal = {IEEE Open J. Commun. Soc.},
  title   = {Power Allocation Optimization for Secure {OFDM}-{NOMA} Downlink Systems},
  year    = {2025},
  month   = {Sept.},
  volume  = {6},
  pages   = {7555--7566},
  doi     = {10.1109/OJCOMS.2025.3610253}
}

@article{Nguyen2025Reliable,
  author={Nguyen, Ti Ti and Ha, Vu Nguyen and Le, Thanh-Dung and Tran, Duc-Dung and Chatzinotas, Symeon and Nguyen, Kim-Khoa},
  journal={IEEE trans. mobile comput.}, 
  title={Reliable Intelligent Reflecting Surface-Assisted Mobile Edge Computing Systems: A Physical Layer Security and Encryption Design}, 
  month = {Feb.},
  year={2026},
  volume={25},
  number={2},
  pages={1950-1966},
  doi={10.1109/TMC.2025.3607599}}

@article{He2019Joint,
  author  = {H. He and P. Ren},
  journal = {IEEE Wireless Commun. Lett.},
  title   = {Joint Artificial Noise and Repetition Coding for Secure Wireless Communications in {TDD} Systems},
  month   = {Dec.},
  year    = {2019},
  volume  = {8},
  number  = {6},
  pages   = {1700--1703},
  doi     = {10.1109/LWC.2019.2937859}
}

@article{Zhao2025Joint,
  author  = {S. Zhao and X. Zhu and Y. Zhang and Z. Zhang and Y. Shen},
  journal = {IEEE Trans. Mobile Comput.},
  title   = {Joint {RIS} and Beamforming Design for Secure and Energy-Efficient Two-Way Relay Communications},
  month   = {Aug.},
  year    = {2025},
  volume  = {24},
  number  = {8},
  pages   = {7440--7457},
  doi     = {10.1109/TMC.2025.3549445}
}

@article{Niu2025Survey,
  author={Niu, Hong and Xiao, Yue and Lei, Xia and Chen, Jiangong and Xiao, Zhihan and Li, Mao and Yuen, Chau},
  journal={IEEE commun. surveys tuts.}, 
  title={A Survey on Artificial Noise for Physical Layer Security: Opportunities, Technologies, Guidelines, Advances, and Trends}, 
  year={2026},
  volume={28},
  number={},
  pages={341-381},
  doi={10.1109/COMST.2025.3610758}}

@article{Fang2022Intelligent,
  author  = {S. Fang and G. Chen and Z. Abdullah and Y. Li},
  journal = {IEEE Commun. Lett.},
  title   = {Intelligent Omni Surface-Assisted Secure {MIMO} Communication Networks With Artificial Noise},
  year    = {2022},
  month   = {Jun.},
  volume  = {26},
  number  = {6},
  pages   = {1231--1235},
  doi     = {10.1109/LCOMM.2022.3159575}
}

@article{Yang2020Artificial,
  author  = {P. Yang and X. Qiu and F. Mu},
  journal = {IEEE Commun. Lett.},
  title   = {Artificial Noise-Aided Secure Generalized Spatial Modulation for Multiuser Transmission},
  year    = {2020},
  month   = {Nov.},
  volume  = {24},
  number  = {11},
  pages   = {2416--2420},
  doi     = {10.1109/LCOMM.2020.3011284}
}

@article{Yi2025Secrecy,
  author  = {Y. Yi and X. Hu and C. Kai},
  journal = {IEEE Commun. Lett.},
  title   = {Secrecy Energy Efficiency Maximization for {ARIS}-Assisted Multiuser {MISO} Systems},
  year    = {2025},
  month   = {Mar.},
  volume  = {29},
  number  = {3},
  pages   = {467--471},
  doi     = {10.1109/LCOMM.2024.3524559}
}

@article{Luo2024Joint,
  author  = {H. Luo and Q. Li and Q. Zhang},
  journal = {IEEE Trans. Inf. Forensics Secur.},
  title   = {Joint Secure Beamforming and Power Splitting Design for {MIMO} Relay Assisted Over-the-Air Computation Networks With Imperfect {CSI}},
  year    = {2024},
  month   = {Jul.},
  volume  = {19},
  pages   = {7075--7090},
  doi     = {10.1109/TIFS.2024.3430549}
}

@article{Chen2024NFDM,
  author  = {J. Chen and Y. Xiao and K. Liu and Y. Zhong and X. Lei and M. Xiao},
  journal = {IEEE Trans. Veh. Technol.},
  title   = {Physical Layer Security for Near-Field Communications via Directional Modulation},
  year    = {2024},
  month   = {Aug.},
  volume  = {73},
  number  = {8},
  pages   = {12242--12246},
  doi     = {10.1109/TVT.2024.3382324}
}

@article{Nasir2024MaxMin,
  author  = {A. A. Nasir},
  journal = {IEEE Commun. Lett.},
  title   = {Max--Min Secrecy Rate Optimization Through Beam Focusing in Near-Field Communications},
  year    = {2024},
  month   = {Jul.},
  volume  = {28},
  number  = {7},
  pages   = {1594--1598},
  doi     = {10.1109/LCOMM.2024.3406396}
}

@article{Zhang2025NFPLS,
  author={Zhang, Yunpu and Fang, Yuan and You, Changsheng and Angela Zhang, Ying-Jun and Cheung So, Hing},
  journal={IEEE Trans. Commun.}, 
  title={Performance Analysis and Low-Complexity Beamforming Design for Near-Field Physical Layer Security}, 
  year={2026},
  volume={74},
  number={},
  pages={781-796},
  doi={10.1109/TCOMM.2025.3626021}}

@article{Tang2025ANBF,
  author  = {Z. Tang and N. Yang and X. Zhou and S. Durrani and M. Juntti and J. M. Jornet},
  journal = {IEEE Trans. Veh. Technol.},
  title   = {Low-Complexity Artificial Noise-Aided Beam Focusing Design in Near-Field {THz} Communications},
  year    = {2025},
  month   = {early access,},
  note    = {\href{https://doi.org/10.1109/TVT.2025.3625557}{doi: 10.1109/TVT.2025.3625557}}
}

@article{Petrov2025WFH,
  author  = {V. Petrov and H. Guerboukha and A. Singh and J. M. Jornet},
  journal = {IEEE Trans. Commun.},
  title   = {Wavefront Hopping for Physical Layer Security in {6G} and Beyond Near-Field {THz} Communications},
  year    = {2025},
  month   = {May},
  volume  = {73},
  number  = {5},
  pages   = {2998--3011}
}

@article{FAS_Secure_Covert,
  author  = {J. Yao and L. Xin and T. Wu and M. Jin and K.-K. Wong and C. Yuen and H. Shin},
  journal = {IEEE Internet Things J.},
  title   = {{FAS} for Secure and Covert Communications},
  month   = {Jun.},
  year    = {2025},
  volume  = {12},
  number  = {11},
  pages   = {18414-18418},
  doi     = {10.1109/JIOT.2025.3539737}
}

@article{Liu2025IM_CT_FAS,
   author={Liu, Min and Xiao, Yue and Zhang, Lechen and Yang, Shuaixin and Wu, Chaowu and Lei, Xia},
  journal={IEEE trans. veh. technol.}, 
  title={Index Modulation for Covert Transmission in Continuous-Trajectory Fluid Antenna Systems}, 
  year={2026},
  month={Mar.},
  volume={75},
  number={3},
  pages={5197-5202},
  doi={10.1109/TVT.2025.3612838}}

@article{FAS_Secrecy_Analysis,
  author  = {F. R. Ghadi and K.-K. Wong and F. J. L{\'o}pez-Mart{\'i}nez and W. K. New and H. Xu and C.-B. Chae},
  journal = {IEEE Trans. Wireless Commun.},
  title   = {Physical Layer Security Over Fluid Antenna Systems: Secrecy Performance Analysis},
  year    = {2024},
  month   = {Dec.},
  volume  = {23},
  number  = {12},
  pages   = {18201--18213},
  doi     = {10.1109/TWC.2024.3463488}
}

@inproceedings{RIS_FAS_Secrecy,
  author    = {F. R. Ghadi and K.-K. Wong and M. Kaveh and F. J. L{\'o}pez-Mart{\'i}nez and W. K. New and H. Xu},
  booktitle = {Proc. IEEE Wireless Commun. Netw. Conf. (WCNC)},
  title     = {Secrecy Performance Analysis of {RIS}-Aided Fluid Antenna Systems},
  year      = {2025},
  month     = {Mar.},
  pages     = {1--6},
  address   = {Milan, Italy}
}

@article{CUMA_Compact_Ultra_Massive,
  author  = {J. D. Vega S{\'a}nchez and H. R. Carvajal Mora and N. V. Orozco Garz{\'o}n and F. J. L{\'o}pez-Mart{\'i}nez},
  journal = {IEEE Open J. Commun. Soc.},
  title   = {Reliable and Secure Communications Through Compact Ultra-Massive Antenna Arrays},
  year    = {2024},
  month   = {Nov.},
  volume  = {5},
  pages   = {7641-7652},
  doi     = {10.1109/OJCOMS.2024.3508463}
}

@inproceedings{Trajectory_FA_Secure,
  author    = {B. Feng and Y. Wu},
  booktitle = {Proc. IEEE 101st Veh. Technol. Conf. (VTC2025-Spring)},
  title     = {Antenna Trajectory-Aware Mechanical Fluid Antenna-Enhanced Secure Wireless Communications},
  year      = {2025},
  month     = {Jun.},
  pages     = {1--6},
  address   = {Oslo, Norway}
}

@article{Chen2025FA_3D_Covert_Jamming,
  author={Chen, Weiyu and Luo, Junshan and Ding, Haiyang and Wang, Shilian and Gong, Fengkui},
  journal={IEEE Trans. Commun.}, 
  title={Three-Dimensional Fluid Antenna-Assisted Covert Communications With Friendly Jamming}, 
  year={2026},
  volume={25},
  number={},
  pages={4156-4170},
  doi={10.1109/TWC.2025.3609276}}

@article{Su2024Sensing,
  author  = {N. Su and F. Liu and C. Masouros},
  journal = {IEEE Trans. Wireless Commun.},
  title   = {Sensing-Assisted Eavesdropper Estimation: An {ISAC} Breakthrough in Physical Layer Security},
  year    = {2024},
  month   = {Apr.},
  volume  = {23},
  number  = {4},
  pages   = {3162--3174},
  doi     = {10.1109/TWC.2023.3306029}
}

@article{Hou2024Optimal,
  author  = {K. Hou and S. Zhang},
  journal = {IEEE J. Sel. Areas Commun.},
  title   = {Optimal Beamforming for Secure Integrated Sensing and Communication Exploiting Target Location Distribution},
  year    = {2024},
  month   = {Nov.},
  volume  = {42},
  number  = {11},
  pages   = {3125--3139},
  doi     = {10.1109/JSAC.2024.3431573}
}

@article{Wu2024Enabling,
  author  = {Z. Wu and M. Cui and L. Dai},
  journal = {IEEE Trans. Wireless Commun.},
  title   = {Enabling More Users to Benefit From Near-Field Communications: From Linear to Circular Array},
  year    = {2024},
  month   = {Apr.},
  volume  = {23},
  number  = {4},
  pages   = {3735--3748},
  doi     = {10.1109/TWC.2023.3310912}
}

@article{Soderi2024MultiRIS,
  author  = {S. Soderi and A. Brighente and S. Xu and M. Conti},
  journal = {IEEE Trans. Mobile Comput.},
  title   = {Multi-{RIS}-Aided {VLC} Physical Layer Security for {6G} Wireless Networks},
  month   = {Dec.},
  year    = {2024},
  volume  = {23},
  number  = {12},
  pages   = {15182--15195},
  doi     = {10.1109/TMC.2024.3452963}
}

@article{Zhang2025UNISAC,
   author={Zhang, Zhentian and Wong, Kai-Kit and Dang, Jian and Zhang, Zaichen and Chae, Chan-Byoung},
  journal={IEEE Journal on Selected Areas in Communications}, 
  title={On Fundamental Limits for Fluid Antenna-Assisted Integrated Sensing and Communications for Unsourced Random Access}, 
  year={2026},
  volume={44},
  number={},
  pages={136-149},
  doi={10.1109/JSAC.2025.3608113}}

@article{Zhang2025FARDIM,
   author={Zhang, Peng and Dang, Jian and Wen, Miaowen and Zhang, Zaichen and Wu, Liang and Yao, Yudong},
  journal={IEEE J. Sel. Areas Commun.}, 
  title={Fluid Antenna-Assisted Rectangular Differential Index Modulation: A Non-Coherent System Design, Optimization, and Performance Analysis}, 
  year={2026},
  volume={44},
  number={},
  pages={1307-1321},
  doi={10.1109/JSAC.2025.3616068}}

@article{Zhang2025SF_FAMA,
  author  = {Z. Zhang and K.-K. Wong and J. Dang and Z. Zhang and C. Masouros and C.-B. Chae},
  journal = {IEEE Wireless Commun. Lett.},
  title   = {On Fundamental Limits of Slow-Fluid Antenna Multiple Access for Unsourced Random Access},
  year    = {2025},
  month   = {Nov.},
  volume  = {14},
  number  = {11},
  pages   = {3455--3459},
  doi     = {10.1109/LWC.2025.3594112}
}

@article{Zhang2024RDRSM,
  author  = {P. Zhang and X. Jin and C. Wan and S. Xing and C. Huang and M. Wen and Y. Yao},
  journal = {IEEE Trans. Commun.},
  title   = {Rectangular Differential Reflecting Spatial Modulation: A Noncoherent Joint Index Modulation of {RIS}-Assisted {MIMO} System},
  month   = {Dec.},
  year    = {2024},
  volume  = {72},
  number  = {12},
  pages   = {7387--7400},
  doi     = {10.1109/TCOMM.2024.3409538}
}

@article{Zhang2019BeamSequence,
  author  = {D. Zhang and A. Li and M. Shirvanimoghaddam and P. Cheng and Y. Li and B. Vucetic},
  journal = {IEEE Trans. Wireless Commun.},
  title   = {Codebook-Based Training Beam Sequence Design for Millimeter-Wave Tracking Systems},
  volume  = {18},
  number  = {11},
  pages   = {5333--5349},
  month   = {Nov.},
  year    = {2019},
  doi     = {10.1109/TWC.2019.2935731}
}

@article{Ma2023RISCellFree,
  author  = {X. Ma and D. Zhang and M. Xiao and C. Huang and Z. Chen},
  journal = {IEEE Trans. Wireless Commun.},
  title   = {Cooperative Beamforming for {RIS}-Aided Cell-Free Massive {MIMO} Networks},
  volume  = {22},
  number  = {11},
  pages   = {7243--7258},
  month   = {Nov.},
  year    = {2023},
  doi     = {10.1109/TWC.2023.3249241}
}

@ARTICLE{wang2024SensingCovert,
  author={Wang, Xinyi and Fei, Zesong and Liu, Peng and Zhang, J. Andrew and Wu, Qingqing and Wu, Nan},
  journal={IEEE Trans. Wireless Commun.}, 
  title={Sensing-Aided Covert Communications: Turning Interference Into Allies}, 
  month = {Step.},
  year={2024},
  volume={23},
  number={9},
  pages={10726-10739},
  doi={10.1109/TWC.2024.3374775}}

@ARTICLE{Xia2026ISACBeamforming,
  author={Xia, Fanghao and Fei, Zesong and Wang, Xinyi and Su, Nanchi and Wang, Zhaolin and Liu, Yuanwei and Xu, Jie},
  journal={IEEE Trans. Wireless Commun.}, 
  title={Toward Secure {ISAC} Beamforming: How Many Dedicated Sensing Beams Are Required?}, 
  year={2026},
  volume={25},
  number={},
  pages={12868-12882},
  doi={10.1109/TWC.2026.3667704}}

@article{Zhang2026FiniteFA,
  author  = {Z. Zhang and K.-K. Wong and H. Jiang and F. R. Ghadi and H. Shin and Y. Zhang},
  journal = {IEEE Wireless Commun. Lett.},
  title   = {Finite-Aperture Fluid Antenna Array Design: Analysis and Algorithm},
  year    = {2026},
  month   = {early access,},
  note    = {\href{https://doi.org/10.1109/LWC.2026.3685911}{doi: 10.1109/LWC.2026.3685911}}
}

@article{Zhang2026JADCE,
  author  = {Z. Zhang and J. Dang and D. Morales-Jimenez and H. Jiang and Z. Zhang and C. Masouros and C.-B. Chae},
  journal = {IEEE J. Sel. Top. Signal Process.},
  title   = {Joint Activity Detection and Channel Estimation for Fluid Antenna System Exploiting Geographical and Angular Information},
  year    = {2026},
  month   = {early access,},
  note    = {\href{https://doi.org/10.1109/JSTSP.2026.3673148}{doi: 10.1109/JSTSP.2026.3673148}}
}

@ARTICLE{Li2026MA_Mobility,
  author={Li, Kaixuan and Yu, Kan and Ma, Dingyou and Zhao, Yujia and Liu, Xiaowu and Zhang, Qixun and Feng, Zhiyong},
  journal={IEEE Trans. Mobile Comput.}, 
  title={Can Movable Antenna-Enabled Micro-Mobility Replace UAV-Enabled Macro-Mobility? A Physical Layer Security Perspective}, 
  month = {Mar.},
  year={2026},
  volume={25},
  number={3},
  pages={4317-4330},
  doi={10.1109/TMC.2025.3624340}}

\end{document}